\documentclass[twocolumn,
10pt,
amsmath,
amssymb,
nofootinbib,
showpacs,
superscriptaddress,
floatfix
]{revtex4-1}
\usepackage{graphicx}
\usepackage{color}
\usepackage[usenames,dvipsnames]{xcolor}
\usepackage[colorlinks=true,
linkcolor=blue,
citecolor=blue,
linktoc=page]{hyperref}
\usepackage{multirow}
\usepackage{float}
\usepackage{flushend}
\usepackage{balance}
\usepackage[varg]{txfonts}
\usepackage{ulem}
\usepackage{fancyhdr}
\usepackage{braket}
\usepackage{mathtools}
\usepackage{lipsum}

\begin{document}

\title{Thermoelectric current in a graphene Cooper pair splitter}

\author{Z. B. Tan}
\affiliation{Low Temperature Laboratory, Department of Applied Physics, Aalto University, Espoo, Finland}
\affiliation{Shenzhen Institute for Quantum Science and Engineering, and Department of Physics, Southern University of Science and Technology, Shenzhen 518055, China}
\author{A. Laitinen}
\affiliation{Low Temperature Laboratory, Department of Applied Physics, Aalto University, Espoo, Finland}
\author{N. S. Kirsanov}
\affiliation{Terra Quantum AG, St.\,Gallerstrasse 16A, 9400 Rorschach, Switzerland}
\affiliation{Moscow Institute of Physics and Technology, 141700, Institutskii
Per. 9, Dolgoprudny, Moscow Distr., Russian Federation}
\affiliation{Low Temperature Laboratory, Department of Applied Physics, Aalto University, Espoo, Finland}
\affiliation{Consortium for Advanced Science and Engineering (CASE), University of Chicago, 5801 S Ellis Ave, Chicago, IL 60637, USA}
\author{A. Galda}
\affiliation{James Franck Institute, University of Chicago, Chicago, IL 60637, USA.}
\affiliation{Materials Science Division, Argonne
National Laboratory, 9700 S. Cass Ave., Argonne, IL 60439, USA}
\author{V. M. Vinokur}
\affiliation{Materials Science Division, Argonne
National Laboratory, 9700 S. Cass Ave., Argonne, IL 60439, USA}
\affiliation{Consortium for Advanced Science and Engineering (CASE), University of Chicago, 5801 S Ellis Ave, Chicago, IL 60637, USA}
\author{M. Haque}
\affiliation{Low Temperature Laboratory, Department of Applied Physics, Aalto University, Espoo, Finland}
\author{A. Savin}
\affiliation{Low Temperature Laboratory, Department of Applied Physics, Aalto University, Espoo, Finland}
\author{D. S. Golubev}
\affiliation{QTF Centre of Excellence, Department of Applied Physics, Aalto University, FI-00076 Aalto, Finland}
\author{G. B. Lesovik}
\affiliation{Terra Quantum AG, St.\,Gallerstrasse 16A, 9400 Rorschach, Switzerland}
\affiliation{Moscow Institute of Physics and Technology, 141700, Institutskii
Per. 9, Dolgoprudny, Moscow Distr., Russian Federation}
\author{P. J. Hakonen}\email[Corresponding author: pertti.hakonen@aalto.fi]{}
\affiliation{Low Temperature Laboratory, Department of Applied Physics, Aalto University, Espoo, Finland}
\affiliation{QTF Centre of Excellence, Department of Applied Physics, Aalto University, FI-00076 Aalto, Finland}

\begin{abstract}
Thermoelectric effect generating electricity from thermal gradient and vice versa appears in numerous generic applications. 
Recently, 
an original prospect of thermoelectricity arising from the nonlocal Cooper pair splitting\,\cite{Lesovik2001,Recher2001} (CPS) and the elastic co-tunneling (EC) in hybrid normal metal-superconductor-normal metal (NSN) structures was foreseen\,\,\cite{Sanchez2018,Kirsanov2019,Hussein2019}. 
Here we demonstrate experimentally the existence of 
non-local Seebeck effect in a graphene-based CPS device comprising two quantum dots connected to an aluminum superconductor 
and theoretically 
validate the observations. 
This non-local Seebeck effect offers an efficient tool for producing entangled electrons.
\end{abstract}

%\pacs{}
\date{\today}
\maketitle
Mesoscopic thermoelectric effects have been observed in a variety of condensed matter systems, including quantum dots\,\cite{Staring1993,Godijn1999,Small2003,Llaguno2004,Scheibner2005}, atomic point contacts\,\cite{Ludoph1999,Reddy2007,Widawsky2012}, Andreev interferometers\,\cite{Eom2005,Jiang2005}, and nanowire heat engines\,\cite{Linke18}. 
Thermoelectric effects in the superconducting systems\,\cite{Hofer1,Hofer2}, especially exploring non-local thermoelectric currents in superconductor-ferromagnet devices\,\cite{Machon} and in bulk non-magnetic hybrid NSN structures\,\cite{Tero,KZ2,KZ3} have attracted special attention. 
Connection between thermoelectric effects and the CPS, proposed in Ref.\,\onlinecite{Cao2015}, established a new mechanism for the coherent non-local thermoelectric effect in hybrid superconducting systems. 
This connection was further studied and explicitly described for a ballistic NSN structure\,\cite{Kirsanov2019}. 
A related nontrivial phenomenon, revealed analytically in Ref.\,\onlinecite{Kirsanov2019}, was that contrary to the intuitive expectations, the superconductor can mediate transfer of heat.
Furthermore, it has been shown that non-local processes depend directly on the EC and CPS probabilities which, in turn, can be made energy-dependent by incorporating quantum dots between each normal lead and the superconducting region\,\cite{Recher2001}. 
Here we present experimental observation of the non-local thermoelectric current generated by imposing thermal gradient across a quantum dot -- superconductor -- quantum dot (QD-S-QD) splitter. 
We find that both CPS and EC processes contribute to the non-local thermoelectric current, and that their relative contributions can be tuned by gate potentials. The ability to tune between CPS and EC allows for testing of fundamental theoretical concepts relating entanglement and heat transport in the graphene CPS systems.
\begin{figure}
\centering
	\includegraphics[width=\linewidth]{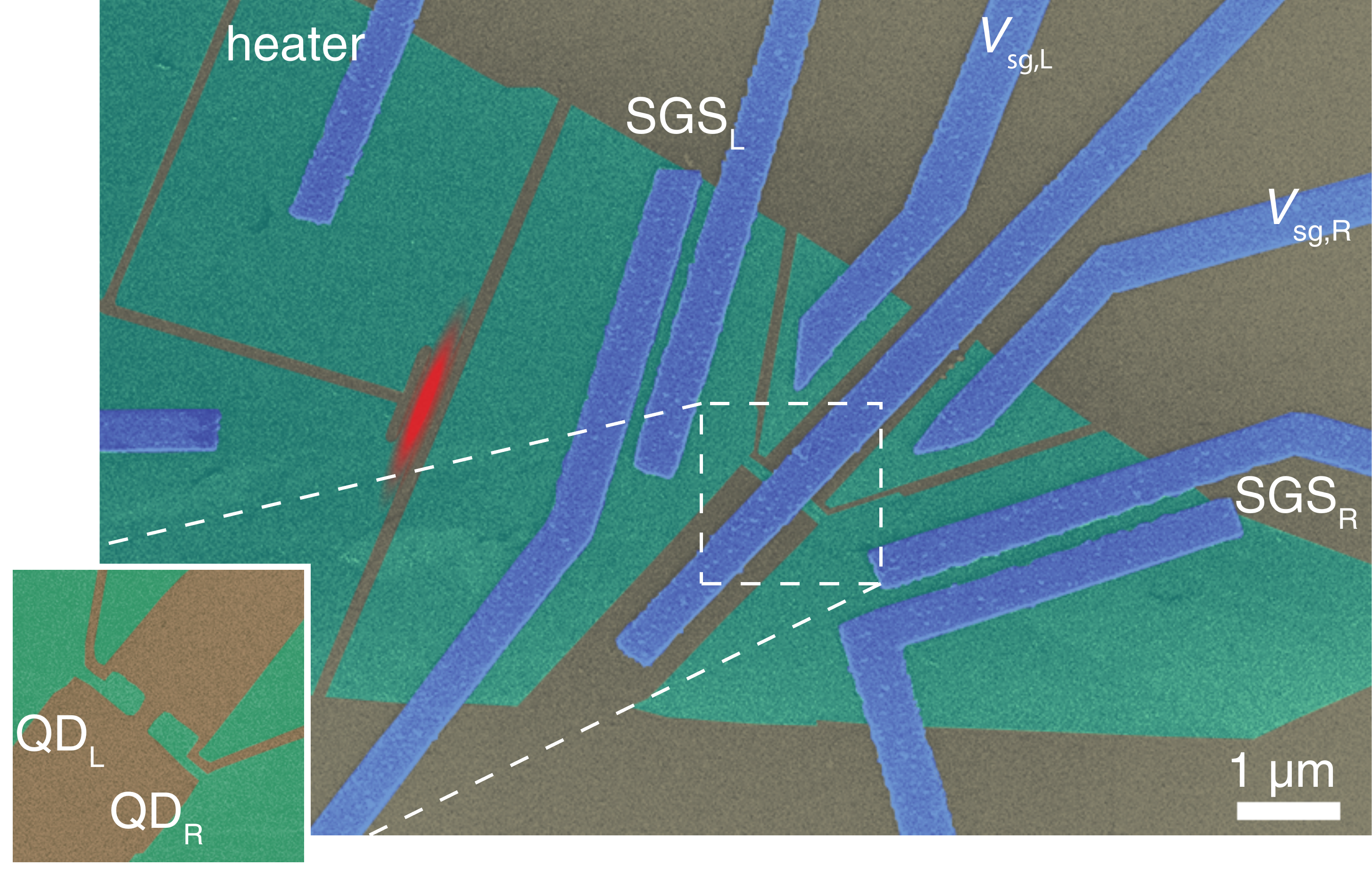}
	\caption{	\textbf{False color SEM image of the device}: green indicates graphene, blue corresponds to metallic Al/Ti sandwich leads, and the substrate background is colored in gray. The Joule heated region is indicated by red color. The superconducting graphene junctions are located between the leads marked by $SGS_{L}$ and $SGS_{R}$. The left and right graphene quantum dots $QD_{L}$ and $QD_{R}$, respectively, have an area $200 \times 150$ nm$^2$, foremost located under the Al injector and thus invisible in the image. Side gates with voltages $V_{sg,L}$ and $V_{sg,R}$ are also carved out of graphene. The inset at lower left corner illustrates the graphene quantum dots before overlaying the metallic Cooper pair injector. }
	\label{fig:SEM}
\end{figure}
\begin{figure*}
\noindent\centering{\includegraphics[width=\linewidth]{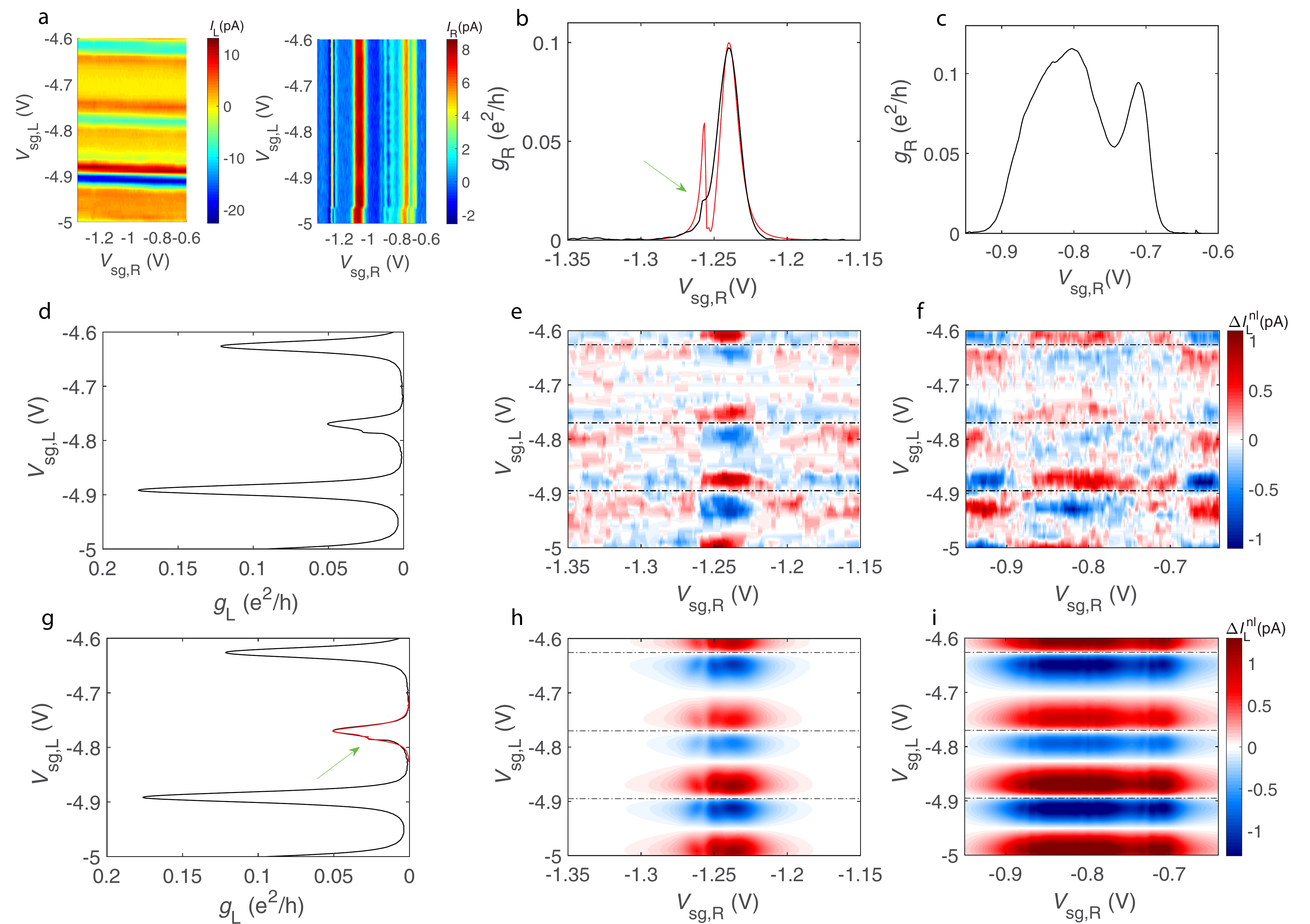}}
\caption{\textbf{Local and non-local contributions to the thermoelectric current. a,} Thermally-generated current at $2f$ in the left and right dot measured as a function of $V_{sg,L}$ and $V_{sg,R}$; the data includes both local and non-local thermoelectric contributions. \textbf{(b,c),} Zero bias conductance of the right quantum dot vs. $V_{sg,R}$ in two intervals: $-1.35\;{\rm V} < V_{sg,R} < -1.15$\,V,
and $-0.95\;{\rm V} < V_{sg,R} < -0.6$\,V; green arrow in the panel b points to the minor peak in the vicinity of the main conductance peak,
and the red curve is the fit by the Fano resonance model with the parameters
$\Gamma_R=20$ $\mu$eV, $\gamma_R=252$ $\mu$eV, $\varepsilon_R-\varepsilon'_R=120$ $\mu$eV, $t_R=55$ $\mu$eV; the additional Fano peak in the fit is intentionally made stronger than the experimentally observed one in order to better reproduce the behavior of the non-local
thermal current in Fig. \ref{exp}d.
\textbf{d,} Zero bias conductance of the left quantum dot in the interval $-5\;{\rm V} < V_{sg,L} < -4.6$ V.
\textbf{(e,f),} Experimental non-local contribution to the termal current in the left quantum dot, $\Delta I^{\rm nl}_L$, in two selected regions of $(V_{sg,R},V_{sg,L})$ plane.
\textbf{g,}  Zero bias conductance of the left quantum dot; data as in panel d, but the red curve displays the fit of one of the peaks with the Fano
resonance model with the parameters $\Gamma_L=6$ $\mu$eV, $\gamma_L=98$ $\mu$eV,
$\varepsilon_L-\varepsilon'_L=24$ $\mu$eV and $t_L=10$ $\mu$eV.
\textbf{(h,i),} Theoretically predicted non-local contribution $\Delta I^{\rm nl}_L$ in the same regions as in panels e and f. }
\label{fig:condarms}
\end{figure*}

We begin with considering an QD-S-QD device within the Landauer formalism.
Taking that the non-local transport is primarily coherent and that the electron energies are smaller than the superconducting gap,  $|E|<\Delta$, 
we find, see Supplementary Information (SI) Note 4, that the EC, $\tau_{\rm EC}(E)$, and CPS, $\tau_{\rm CPS}(E)$, probabilities are given by the expressions
\begin{eqnarray}
\tau_{\rm EC} = \tau_L(E) \tau_S \tau_R(E),\; \tau_{\rm CPS} = \tau_L(E) \tau_S \tau_R(-E).
\label{tau}
\end{eqnarray}
Here $\tau_{L(R)}(E)$ is the transmission probability of the left (right) quantum dot which depends on the energy of an electron and
on the side gate potentials applied to the dots $V_{sg,L(R)}$
($\tau_{L(R)}(E)$ is given by the sum of Lorentzian peaks or Fano resonances 
associated with discrete energy levels (SI Note 4) and $\tau_S$ is the effective transmission probability of the superconducting lead.
The latter corresponds to the probability for an electron coming out of one dot so that, instead of escaping into the bulk of the superconducting electrode, it reaches the other dot. 
It  becomes independent on the electron energy $E$ if the dots are separated by a distance shorter than the superconducting coherence length. 
This condition is reasonably well fulfilled in our experiment.
The non-local thermoelectric currents in the dots can, in turn, be expressed in terms of the elastic cotunneling and Cooper pair splitting contributions,
$\Delta I^{\rm nl}_L = (\Delta I_{\rm EC} + \Delta I_{\rm CPS})/2$, $\Delta I^{\rm nl}_R = (-\Delta I_{\rm EC} + \Delta I_{\rm CPS})/2$,
where
\begin{eqnarray}
\Delta I_{\rm EC} &=& \frac{2e}{h}\int dE \,\tau_{\rm EC}(E) [f_L(E)-f_R(E)],
\nonumber\\
\Delta I_{\rm CPS} &=& \frac{2e}{h}\int dE \,\tau_{\rm CPS}(E) [f_L(E)-f_R(E)],
\label{IEC}
\end{eqnarray}
and $f_{L(R)}(E)=1/(1+e^{E/k_BT_{L(R)}})$ is the distribution function in the left (right) electrode having the tempertature $T_{L(R)}$.

Now, we turn to the experiment. Our device presented in Fig.~\ref{fig:SEM} consists of an Al superconducting injector in contact with two graphene quantum dots.
Two side gate electrodes allow us to tune the resonance levels of the dots independently. 
In order to perform thermoelectric measurements, our device additionally contains two thermometers and a resistive heater, fabricated from a graphene monolayer. 
The thermometers are superconductor-graphene-superconductor (SGS) Josephson junctions that reveal local temperature through the temperature dependence of the switching current, $I_{\rm sw}(T)$ \cite{Voutilainen2011}. 
The resistive heater comprises the graphene nanoribbon and two attached aluminum leads. 
The heater is distinctly apart and electrically isolated from the rest of the device, the heat to the Cooper pair splitter being transmitted through the substrate.
\begin{figure}
\noindent\centering{\includegraphics[width=\linewidth]{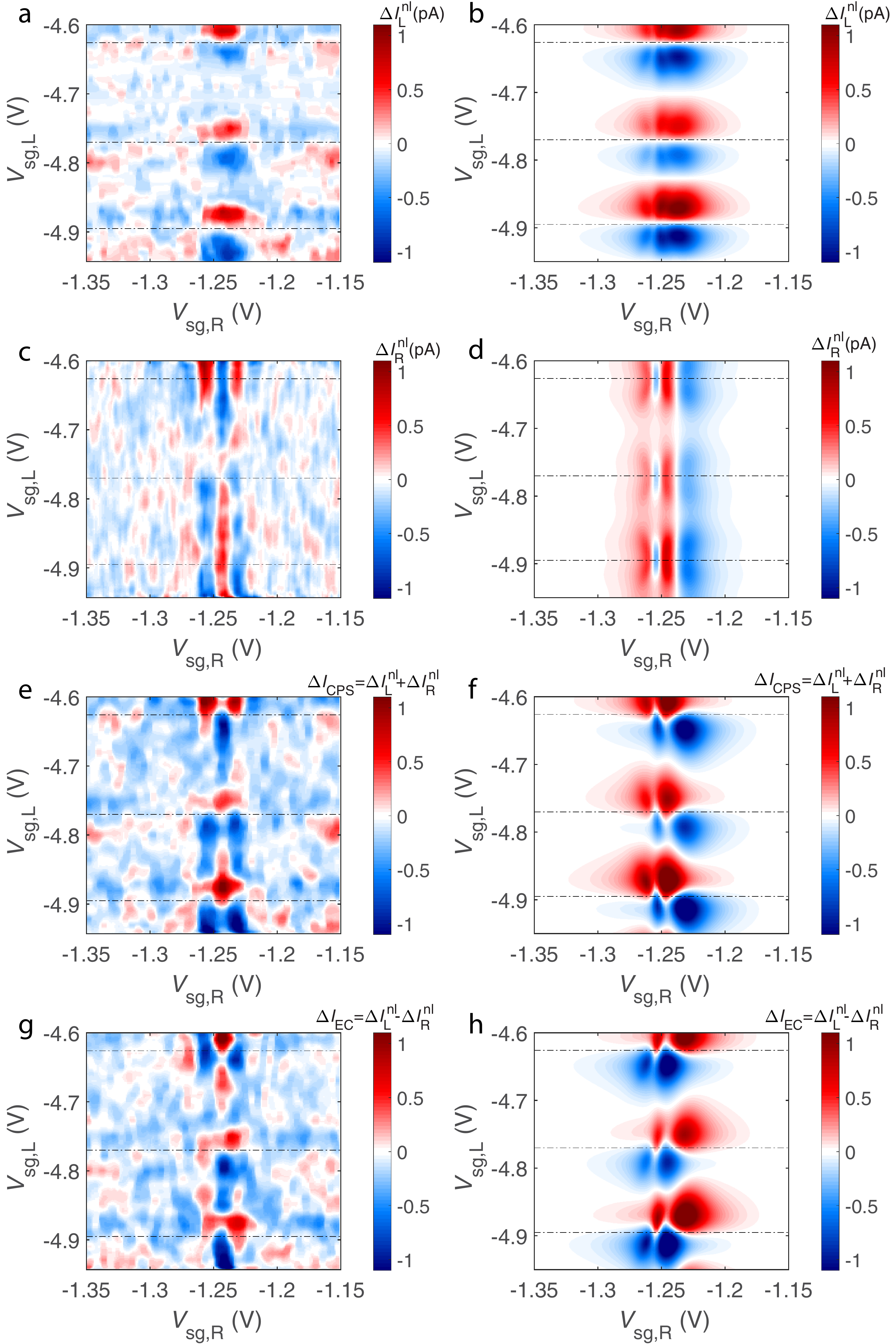}}
	\caption{\textbf{Interplay of non-local thermal  currents in the left and right dots.} Non-local contributions to the thermal currents of the right and left dots \textbf{(a-d)}
     and the corresponding CPS and EC currents \textbf{(e-h)}. The left column (graphs \textbf{a,c,e,g}) shows the experiment and the right one
     (graphs \textbf{b,d,f,h}) -- theory.
     Horizontal dotted lines in the plots indicate the positions of the conductance peak maxima of the left dot.
     %they are slightly shifted from the corresponding positions on the experimental plots due to time instability of the gate potentials. 
     }
	\label{exp}
\end{figure}

The temperature difference $\Delta T=T_L-T_R$ between the leads of the two-terminal device induces the thermoelectric current $I=\alpha G \Delta T$,
where $G$ is the conductance of the device and $\alpha$ is the Seebeck coefficient\,\cite{TeroBook}.
For typical metals, such as aluminum or copper, the Seebeck coefficient is quite small, $\alpha \sim 3-7$\,$\mu$V/K.
For graphene, $\alpha$ is inversely proportional to the square root of charge density, and it can reach much higher values close to the charge neutrality point \cite{Zuev2009,Wei2009}.
In quantum dots with energy-dependent electron transmission probability, large $\alpha$ up to a few $k_B/e\sim 100$\,$\mu$V/K can also be achieved
\cite{Linke}. 
In our experiment, we observe similar values of the Seebeck coefficient in graphene quantum dots.
We operate the graphene heater at frequency $f =2.1$\,Hz and record thermoelectric currents through both quantum dots at the double frequency $2f$ (see Methods). 
Thermal gradient induced by the heater is measured by SGS thermometers, which were calibrated separately as discussed in SI Note 2.
%\,\cite{Supplement}.

The thermoelectric current induced by the heater in the left (right) quantum dot is given by the sum of dominating local ($I_{L(R)}^{\rm loc}$) and small non-local contributions ($\Delta I^{\rm nl}_{L(R)}$), $I_{L(R)}=I_{L(R)}^{\rm loc}(V_{sg,{L(R)}})+ \Delta I^{\rm nl}_{L(R)}(V_{sg,L},V_{sg,R})$, see SI Note 4.   
To infer the non-local contribution from the measured current $I_{L(R)}$, we subtract from $I_{L(R)}$ its value averaged over the different gate voltages on the opposite dot, $\langle I_{L(R)}\rangle$:
\begin{equation}
     \Delta I^{\rm nl}_{L(R)} = I_{L(R)} - \langle I_{L(R)}\rangle.
\end{equation}
%
%As the non-local thermoelectric currents were quite weak, this procedure was carried out separately for each energy level (see Methods). 
We thus obtain non-local currents $\Delta {I}^{\rm nl}_{L(R)}$, which have a magnitude of order of $5-10$\% of the total thermoelectric currents.
Figure \ref{fig:condarms} displays the maps of the non-local thermoelectric current in left dot $\Delta I^{\rm nl}_L(V_{sg,L},V_{sg,R})$ measured in the vicinity of the two conductance peaks of the right dot for the heating voltage $V_h=5$ mV.
In Fig.\,\ref{fig:condarms}(b,c,d,g), $g_{L(R)} = hG_{L(R)}/e^2$ is the dimensionless conductance of the left (right) quantum dot.
We find that $\Delta I^{\rm nl}_L$ is symmetric with respect to the centers of the conductance peaks of the right dot, but it changes sign at the maxima of conductance peaks of the left dot. 
Thus, the non-local current $\Delta I^{\rm nl}_L$ approximately follows the same pattern as the product $[dg_L(V_{sg,L})/dV_{sg,L}] g_R(V_{sg,L})$.

Before proceeding to our main result, note that some conductance peaks are split into two closely located peaks (see Figs. \ref{fig:condarms}(b,c)). 
The splitting is explained by the Fano resonant effect, see SI Note 4. 
Namely, we introduce the coupling rates $\Gamma_{j,n}$ and $\gamma_{j,n}$ (here $j=L,R$ enumerates the dots) between the $n$th energy level of the dot (with energy $\varepsilon_{j,n}$) and, respectively, normal and superconducting leads; we also assume that the $n$th level is coupled to a ``dark" energy level, having the energy $\varepsilon'_{j,n}$, via the hopping matrix element $t_{j,n}$.
This results in  the transmission probabilities of the dots $\tau_j = \sum_{n} \gamma_{j,n}\Gamma_{j,n}/[\left(E-\varepsilon_{j,n}-{|t_{j,n}|^2}/{(E-\varepsilon'_{j,n})}\right)^2 + {(\gamma_{j,n}+\Gamma_{j,n})^2}/{4}]$, see SI Note 4.
The conductances $g_L(V_{sg,j}) = 4\tau_j^2(0,V_{sg,j})/[2-\tau_j(0,V_{sg,j})]^2$, as predicted by the theory of Andreev reflection \cite{Beenakker1992},
exhibit splitted peaks for $t_{j,n}\not=0$.

Figure\,\ref{exp} displays the main result of our study. 
There we plot the non-local thermal currents for both quantum dots together with the theory predictions based on Eqs.\,(\ref{tau}) and (\ref{IEC}).
The involved model parameters are chosen in such a way that, besides accounting well for the non-local current, they can also reasonably fit the conductance peaks (see the caption of Fig.\,\ref{fig:condarms}). 
In the experiment, the non-local current $\Delta I^{\rm nl}_R$  changes its sign three times in the vicinity of the conductance peak of the right dot located at $V_{sg,R}=-1.24$\,V.
Although in order to reproduce this behaviour, we had to take hopping amplitude, $t_R$, larger than required by the perfect fit to the conductance peak (see Fig. \ref{fig:condarms}b), this offers a fair cross-check for our description. 
%We have also treated $\tau_S$ as a fitting parameter and find that the value $\tau_S=0.1$ provides the best fit.
One sees that not only the magnitudes of the currents $\Delta I^{\rm nl}_L$ and $\Delta I^{\rm nl}_R$ are in good agreement with the theory, but their symmetric and anti-symmetric combinations $\Delta I_{\rm CPS}=\Delta I^{\rm nl}_L+\Delta I^{\rm nl}_R$ and $\Delta I_{\rm EC}=\Delta I^{\rm nl}_L-\Delta I^{\rm nl}_R$ exhibit the expected gate voltage dependence, providing a convincing support of the nonlocal coherent thermoelectric effect in our device. 
\begin{figure}
\noindent\centering{\includegraphics[width=\linewidth]{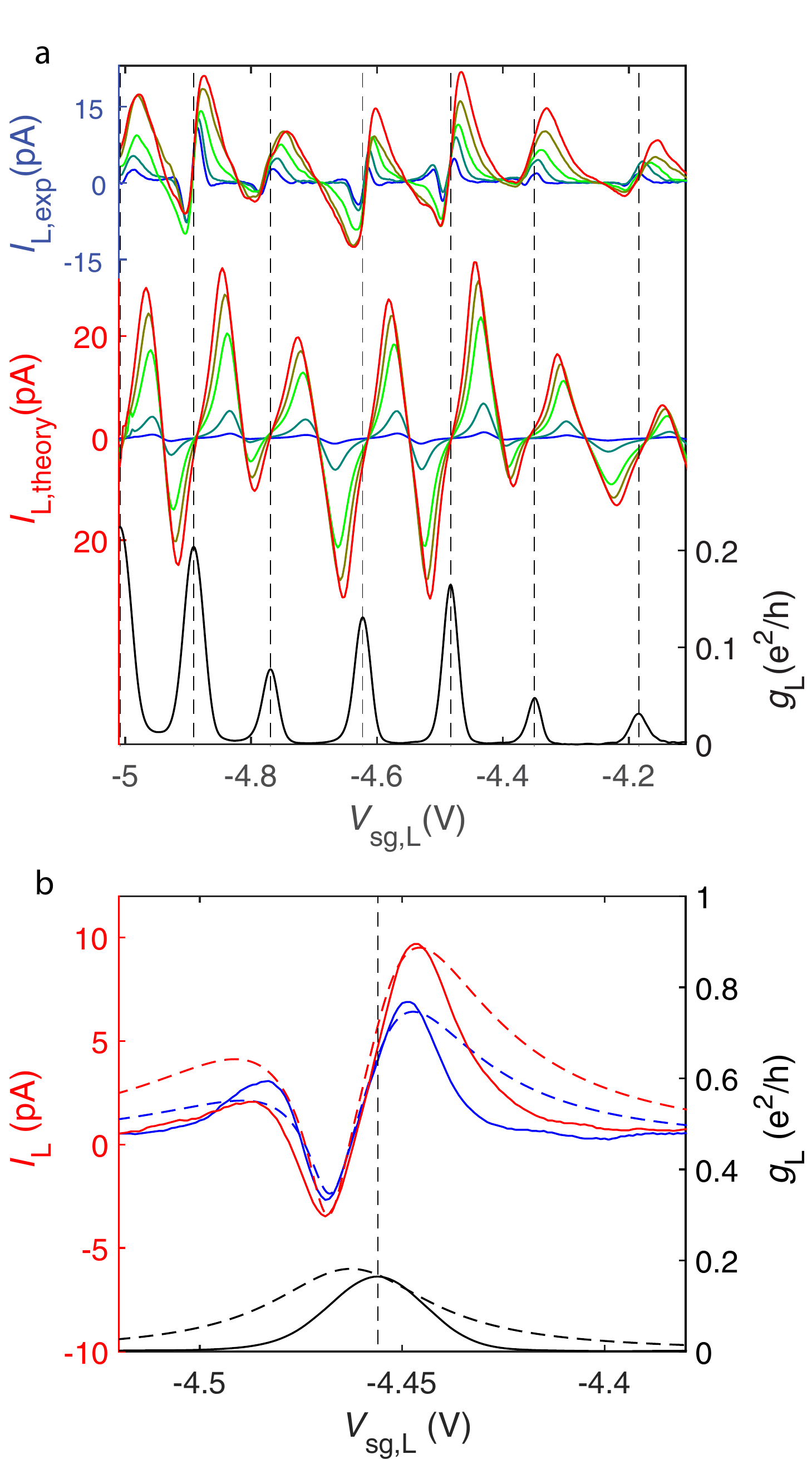}}
\caption{
\textbf{Drive dependence of thermoelectric current. a,} Upper curves: thermal current in the left quantum dot $I_{L,exp}$ \emph{vs}. gate voltage $V_{sg,L}$ measured at heating voltages $V_h=[5, 9, 19, 25, 29] $ mV where $V_h=5 $mV is the blue curve and 29 mV is the red curve. We estimate the induced temperature difference between the left and right quantum dots to be
$T_L-T_R \simeq 17$ mK for $V_h=5$ mV, and
$T_L-T_R \simeq 59 $ mK for $V_h =29 $ mV.
Middle curves: theory predictions based on the coherent model (see SI Note 4) for the thermal current $I_{L,theory}$ plotted in the same manner as the upper curves for $V_h=[5, 10, 20, 25, 30] $ mV. The gap of Al/Ti leads is set to $\Delta_0=150$ $\mu$eV at $T=0$ while the BCS gap formula $\Delta(T_S)$ 
%temperature dependence with the temperature of the superconductor set as 
with $T_S=(T_L+T_R)/2$ defines the $T$ dependence.
Lowest curve: experimental conductance of the left quantum dot.
\textbf{b,} Incoherent modeling for the low temperature regime: theoretical fits (dashed) to the measured thermoelectric currents $I_L$ and conductance $g_L$ (solid) at $V_h=7$ and 11 mV, blue and red curves, respectively.
The model is based on the assumption that the system may be split into the coherent subsystems which, in turn, are joined incoherently
into a circuit.
%The thermal gradient across the dot induces both electric current and voltage drop on the it. 
%We take the temperatures on the left ($T_{LL}$) and right ($T_{LR}$) sides of the left dot equal to $T_{LL}=(0.09+9.1\, V_h^{0.70})$\,K and $T_{LR}=(0.09+8.67\, V_h^{0.70})$\,K.
Details of the model and fitting parameters are given in SI Note 5.}
\label{fig:TEC}
\end{figure}

Next, we discuss local thermoelectricity.
Since, as noted, the non-local currents are relatively small, one can treat the measured currents foremost as local, $I_{L(R)}\simeq I_{L(R)}^{\rm loc}$.
The measured thermoelectric current of the left quantum dot is shown in Fig.\,\ref{fig:TEC}a. 
The lowest curve in this panel shows the dimensionless conductance of the left quantum dot $g_L = hG_L/e^2$ as a function of the side gate voltage $V_{sg,L}$. 
Thermoelectric current $I_L$, depicted by the upper curves of Fig.\,\ref{fig:TEC}a, varies with the same period as the conductance. 
Its magnitude grows with the increasing heating power $P$ as  $I_L^{\max}\propto P^{1/3}$, which is consistent with $G_{\rm th}\propto T^3$ for the thermal conductance between electrons in graphene and phonons in the substrate. 
The maximum thermal power of the left quantum dot reaches a value of $\alpha_{\max} = \max\{I_L/G_L(T_L-T_S)\} \approx 250$ $\mu$V/K, which is close to the values reported in Ref. \onlinecite{Linke}.
Since we cannot reliably measure the temperature of the superconductor $T_S$, we set $T_S=(T_L+T_R)/2$ in evaluating $\alpha_{\max}$ and in our theory modeling.
In Fig.\,\ref{fig:TEC}a we also show the local thermal current predicted by the theory of Andreev reflection with energy-dependent transmission probability\,\cite{Sun1999}; the same theory
was earlier employed in deriving the non-local contributions using Eq. (\ref{IEC}). 
In the local case, only those quasiparticles with energies above the superconducting 
gap, $|E|>\Delta$, contribute. 
The zero temperature value of the gap $\Delta_0$ is set by the transition temperature $T_c=1.0$\,K of the Al/Ti leads, and the transmission probability of the dot $\tau_L(E,V_{sg,L})$  is inferred from the experimentally measured conductance $g_L(V_{sg,L})$, as explained in SI Note 5. We find rather good agreement between theory and experiment except for the lowest values of the heating voltage.
This agreement provides further confirmation for our model.

In the low temperature regime, the coherent model predicts very small current due to lack of quasiparticles, while the experimental thermal current remains significant and exhibits additional sign changes in the vicinity of some of the conductance peaks. 
These features can originate from non-zero, thermally-induced voltages across the dots.
To capture these effects, we propose that electrons may undergo quick inelastic relaxation, see SI Note 5.
This introduces incoherent effects that facilitate description of  quantum dots and NS interfaces as independent conductor elements connected in series.
The results of such an inelastic model are shown in Fig.\,\ref{fig:TEC}b.
The incoherent description accurately predicts the character of the local thermoelectricity at small temperatures. 
Incidentally, although at odds with the effect of local thermoelectricity, the non-local currents are dominantly determined by coherent electrical transport.

This foundational work has demonstrated the use of thermal gradient as \textit{primus motor} for generating entangled electrons in graphene Cooper pair splitter. As the quantum dots in the device can be tuned individually, we are able to tune the device operation between EC and CPS regimes, thereby accomplishing direct control of two streams of entangled electrons. This type of scheme is useful not only for enabling devices where electrical drive is neither possible nor desired, but also as a platform for realizing quantum thermodynamical experiments.

\section*{Methods}
\subsection{Samples and fabrication}
Our graphene films were manufactured using mechanical exfoliation of graphite (Graphenium, NGS Naturgraphit GmbH) and placed on a highly p$^{++}$ doped silicon wafer, coated by 280-nm-thick thermal silicon dioxide. The conducting substrate was employed as a backgate for coarse tuning of the graphene quantum dots, while fine tuning was performed by adjusting the side gates. Electron beam lithography (EBL) on PMMA resist was used to pattern a mask for plasma etching of the graphene structures. A second EBL step was carried out to expose the pattern for electrode structures, followed by deposition of Ti/Al (5nm/50nm, superconducting $T_c=1.0$\,K) leads using an e-beam evaporator.

The strong p$^{++}$ doping and the interfacial scattering at the Si/SiO$_2$ interface reduce the phonon mean free path in the substrate to one micron range, which facilitates the use of the heat diffusion equation for estimating thermal gradients along the substrate near the graphene ribbon heater and the splitter. Heat transport analysis was done separately for each component involved in the operation of the CPS, as well as a COMSOL simulation, see SI Note 1).

\subsection{Measurement scheme}
Our conductance and thermoelectric current measurements employed regular lock-in techniques at low frequencies.
The galvanically-separated heater was driven at $f=2.1$\,Hz, with an ac voltage amplitude ranging between $1-40$\,mV (for data without galvanic separation, see SI Note 3). Because the resistance $R$ of the graphene ribbon heater was independent of temperature in its regime of operation, the heating power $P=V_h^2/R$ was fully governed by the voltage $V_h$. The heating power oscillated at frequency $2f=4.2$\,Hz, which resulted in thermoelectric currents at 4.2 Hz, recorded using a lock-in time constant of 1 sec. The thermal response time of our device appears to be well below 1\,ms, i.e. much less than a measurement period, so that the thermal response is not suppressed. The use of such a low frequency for the experiments was dictated by the need to eliminate the capacitive coupling between the wires in the measurements.

The local temperature was monitored using two SGS junctions. At low temperature, because of proximity effect, graphene becomes superconducting, with a  supercurrent exponentially proportional to temperature: $\sim \exp(-T/E_{\textrm{Th}})$. Here $E_{\textrm{Th}}=\hbar D/L_{SGS}^2$ stands for the Thouless energy given by the length of the SGS section $L_{SGS}$ and the diffusion constant $D \sim \frac{1}{2} v_F \lambda$, where the Fermi velocity $v_F=8 \times 10^5$\,m/s and $\lambda$ is the charge carrier mean free path of graphene. Using $\lambda \simeq 20$\,nm for graphene on SiO$_2$ and $L_{SGS}=200$ \,nm, we estimate $E_{\textrm{Th}} \simeq 1305$ $\mu$\,eV for our SGS thermometers. This yields for the optimum thermometer sensitivity around $T \simeq 0.3$\,K, with tendency towards saturation below $T < 0.2$\,K. However, the maximum differential resistance at the gap edge still grows with decreasing temperature below $T=0.2$ K. The absence of clear hysteresis of the SGS junctions is assigned to the "built-in  shunts" provided by the surrounding, non-proximitized graphene.
%The local temperature was monitored by two SGS junctions. At low temperature, because of proximity effect, graphene becomes superconducting, with a switching supercurrent exponentially proportional to temperature: $\exp(-T/E_{\textrm{Th}})$. Here $E_{\textrm{Th}}=\hbar D/L_{SGS}^2$ stands for the Thouless energy given by the length of the SGS section $L_{SGS}$ and the diffusion constant $D \sim \frac{1}{2} v_F \lambda$, where the Fermi velocity $v_F=8 \times 10^5$\,m/s and $\lambda$ is the charge carrier mean free path of graphene. Using $\lambda \simeq 20$\,nm for graphene on SiO$_2$ and $L_{SGS}=200$ \,nm, we estimate $E_{\textrm{Th}}=125$ $\mu$\,eV for our SGS thermometers. This yields for the optimum thermometer sensitivity around $T \simeq 0.3$\,K, with tendency to saturation below $T < 0.2$\,K.

%As  temperature is increased, the hysteresis of the IV characteristics of the SGS junctions vanishes leading to large peaks in the differential resistance. 
We have used the amplitude of the differential resistance peak $R_d^{\textrm{max}}$ vs. $T$ to infer the effective local temperature within the graphene sample. The SGS temperature under the voltage bias $V_h$ in the graphene ribbon heater was obtained by direct comparison between $R_d^{\textrm{max}}$ and the heating power to the value of $R_d^{\textrm{max}}$ recorded when varying the cryostat temperature. As detailed in SI Note 2, we obtain the relation $T_{SGS,L}=9.1 \times V_h^{0.70} +90$\,mK between the $SGS_L$ temperature and the heating voltage ($V_h$ in Volts). For the $SGS_R$ thermometer we obtained  an estimate $T_{SGS,R} \simeq 8.4 \times V_h^{0.70} +90$\,mK.

\subsection{Theoretical modeling}
Our theoretical calculations are based on both coherent and incoherent modeling of transport. In coherent modeling, we employ the Landauer approach with Andreev reflection \cite{Blonder1982,Beenakker1992} for calculating the local thermoelectric current; Lorentzian resonance line shapes are employed for transport in the quantum dots \cite{Sun1999,Deacon2015,Gramich2015,Gramich2017}. For the non-local current, we employ a standard crossed Andreev reflection formalism \cite{Feinberg2000,Recher2001,Golubev2009,Golubev2019}. In our incoherent theory, based on scattering matrix formalism \cite{Lesovik2001,Sadovskyy2015,Kirsanov2019}, we also include the influence of internal thermally-generated current sources and their ``back-action" effect owing to the environmental impedance caused by graphene ribbons. The inclusion of the back-action-induced voltage sources makes the incoherent calculation self-consistent. For details of the calculations, we refer to Notes 4 and 5 of SI.

\section*{Acknowledgements}
We are grateful to C. Flindt and P. Burset for  discussions on the role of Coulomb blockade in Cooper pair splitting and to I.\,A.\,Sadovskyy for sharing his numerical codes. This work was supported by Aalto University School of Science Visiting Professor grant to G.B.L., as well as by Academy of Finland Projects No. 290346 (Z.B.T., AF post doc), No. 314448 (BOLOSE), and No. 312295 (CoE, Quantum Technology Finland). The work of A.L. was support by the Vilho, Yrj\"{o} and Kalle Väis\"{a}l\"{a} Foundation of the Finnish Academy of Science and Letters. This work was also supported within the EU Horizon 2020 programme by ERC (QuDeT, No. 670743), and in part by Marie-Curie training network project (OMT, No. 722923), COST Action CA16218 (NANOCOHYBRI), and the European Microkelvin Platform (EMP, No. 824109). The work of N.S.K. and G.B.L. was supported by the Government of the Russian Federation (Agreement No. 05.Y09.21.0018), by the RFBR Grants No. 17-02-00396A and No. 18-02-00642A, Foundation for the Advancement of Theoretical Physics and Mathematics BASIS, the Ministry of Education and Science of the Russian Federation No. 16.7162.2017/8.9. The work of N.S.K and A.G. at the University of Chicago was supported by the NSF grant DMR-1809188.
The work of V.M.V. was supported by the U.S. Department of Energy, Office of 
Science, Basic Energy Sciences, Materials Sciences and Engineering Division.
\bigskip

\section*{Author contribution}
This research, initiated by P.J.H., is an outgrowth of a long term collaboration between G.B.L. and P.J.H. The experimental setting and the employed sample configuration were developed by Z.B.T. and P.J.H. The patterned graphene samples were manufactured by Z.B.T. using Aalto University OtaNano infrastructure. The experiments were carried out at OtaNano LTL infrastructure by Z.B.T. and A.L. who were also responsible for the data analysis. A.S. and M.H. were adjusting and operating the LTL infrasturcture. Theory modeling for coherent transport was performed by D.S.G., N.S.K., and G.B.L. The theory for incoherent transport was foremost developed and analyzed by N.S.K., A.G., V.M.V., and G.B.L. The results and their interpretation were discussed among all the authors. The paper and its Supplementary Information were written by the authors together.

\section*{Competing interests} The authors declare that they have no competing interests.

%% Here is the endmatter stuff: Supplementary Info, etc.
%% Use \item's to separate, default label is "Acknowledgements"

\clearpage
\setcounter{figure}{0} 
\setcounter{equation}{0}
\renewcommand{\theequation}{S\arabic{equation}}
\renewcommand{\thefigure}{S\arabic{figure}}
\setcounter{subsection}{0}
\setcounter{subsubsection}{0}
%\include{suppl_kn_v4}
%\begin{widetext}
\onecolumngrid
\section*{Supplementary Information}
This supplement provides additional details on experimental and theoretical analysis of our results. We begin with thermal modelling of our Cooper pair splitter device and discuss thermal transport in various parts of the sample, starting from the graphene heater and its electron-phonon coupling, down to small scale thermal gradients between the two quantum dots on SiO$_2$ (Note 1). In Note 2, we provide differential resistance data on the SGS superconducting junction thermometers and discuss how we can determine the thermal gradients in our samples using calibrations based on the SGS junctions. Additional thermopower data at finite bias is provided in Note 3 in order to stress the importance of having a galvanically separated heater for inducing well-defined thermoelectric currents. Note 4 contains details of our theoretical models.
We present the theory of thermal transport in a system consisting of two quantum dots coupled to two separate normal leads and one common superconducting lead (cf. Fig. \ref{NDSDN} for the setting). In Note 4 we describe the coherent transport model, which appears to agree well with the experimental results. In Note 5, we discuss the incoherent transport regime. This model helps to account for additional sign changes in the local thermal current at low heating power which are seen in the experiments, but not captured by the coherent model. 

\section*{Note 1: Modeling heat transport in the system}

\begin{figure}[h!]
	\includegraphics[width=0.75\textwidth]{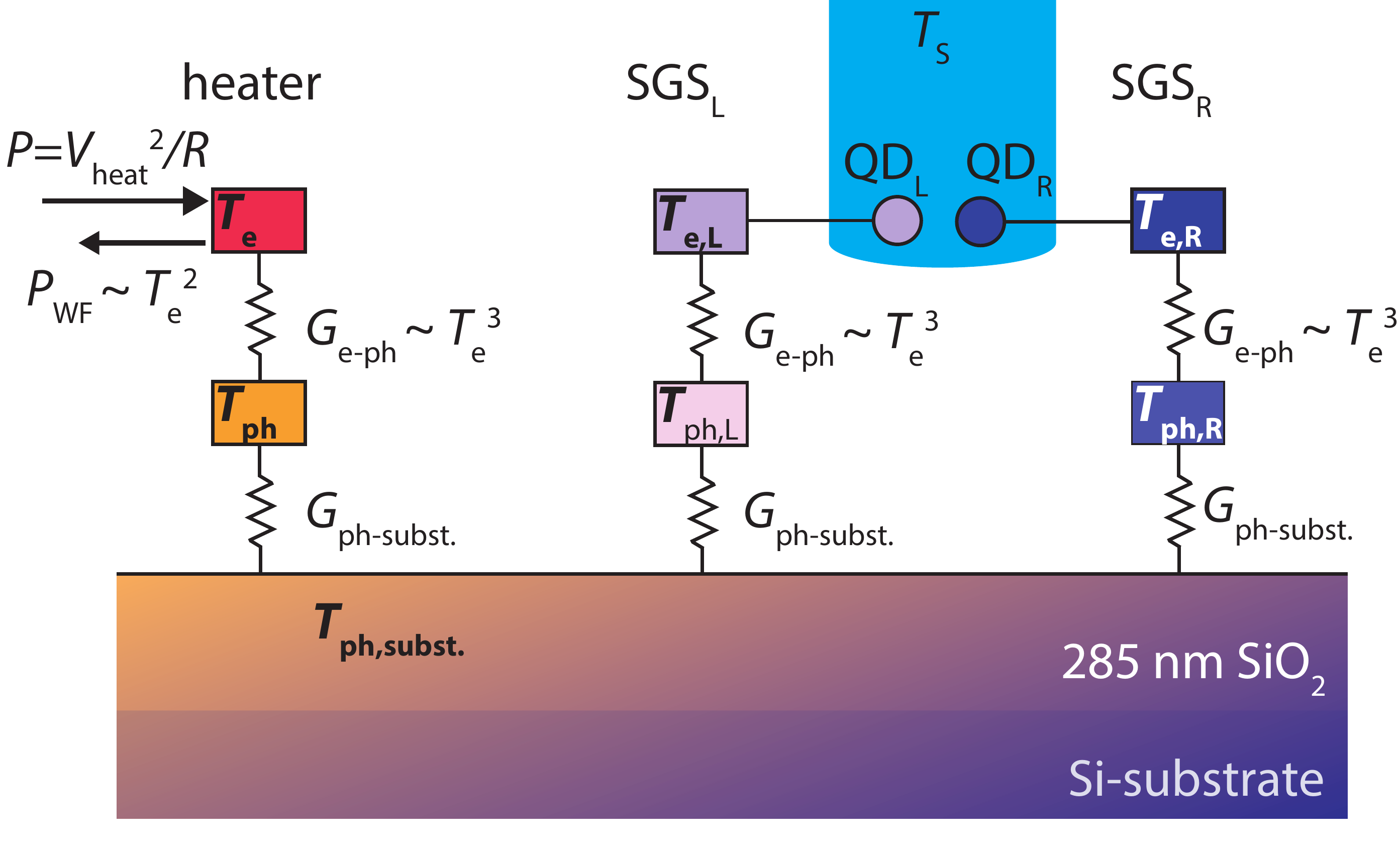}
	\caption{Heat flow model across our 8-$\mu$m-long sample displaying electron and phonon systems in graphene (heater, and superconducting junctions $SGS_L$ and $SGS_R$), as well as phonon system of the substrate. Heat is brought in by Joule heating $P=V_h^2/R_h$ with $R_h \sim 20$ k$\Omega$ at a typical charge carrier density of $n=4 \times 10^{11}$ cm$^{-2}$. The electron-phonon coupling in the graphene ribbon heater is expected to be dominated by impurity-assited acoustic phonon scattering (supercollisions) \cite{Song2012}, while the phonon-phonon heat transport to the substrate is weakened by Kapitza mismatch at the interfaces. Compared with thermal conductivity of the strongly doped Si, the Kapitza resistance between Si and SiO$_2$ can be neglected (The Kapitza resistance corresponds to approximately 4 $\mu$m layer of our doped Si.) For electrically conducting parts, the electrical heat diffusion governed by the Wiedemann-Franz law is included as relevant. Warm (reddish) colors indicate higher temperatures (close to $T_{e,max} = 80$ K), while bluish colors denote lower temperatures (close to the base temperature $T=90$ mK). The total thickness of the sample is 525 microns (not to scale). }
	\label{fig:heatflowdiagram}
\end{figure}

Fig. \ref{fig:heatflowdiagram} displays a schematic heat flow diagram of our graphene Cooper pair splitter device. The heater, the graphene Josephson junctions thermometers ($SGS_L$ and $SGS_R$), and the quantum dots ($QD_L$ and $QD_R$) are all lying on a SiO$_2$/Si substrate. We have left out the Al electrode which connects  thermally $QD_L$ and $QD_R$ owing to quasiparticle resistance. According to theory, the important temperatures for non-local thermoelectric effects are those of the left and right normal reservoirs, $T_L$ and $T_R$ which are formed by the large graphene pieces before the narrow graphene constrictions making the connections to the dots. For the local thermopower also the temperature of the superconductor, $T_S$, is relevant. 
%Therefore, the details of thermal balance in the left-out Al region are of no consequence to our analysis.

Thermal gradient between the two quantum dots is produced by heat flow imposed across the substrate underneath the sample. Establishment of local substrate temperature requires sufficient amount of scattering of phonons, i.e. small enough mean free path for them (see  Sect. \ref{limits}). In our analysis we assume that the local temperatures $T_L$, $T_S$, and $T_R$ can be approximately defined on sub-micron length scales, which is on the order of the separation of the two graphene reservoirs with the SGS thermometers.

The thermal analysis of our system can be divided into four parts: 1) thermal flow in the heater, 2) thermal flow from the heater to the substrate, 3) thermal flow on the SGS thermometer, and 4) thermal flow along the substrate.

\subsection{Thermal flow in graphene heater} \label{LTA}

The simplest model for the temperature rise in the graphene ribbon heater is provided by the hot electron model.  This allows us to estimate the electronic temperature of the graphene electrons using the formula: $T_e=\frac{\sqrt{3}}{8} \frac{eV}{k_B}$, which yields $T_e^{\textrm{max}} \sim 80$ K at the highest employed heating voltages. This temperature represents an average over the heater electrons and the actual temperature distribution over the wire is approximately  quadratic with the end temperatures fixed according to the heat balance in the graphene reservoirs of the ribbon.

\begin{figure}[!ht]
	\includegraphics[width=10cm]{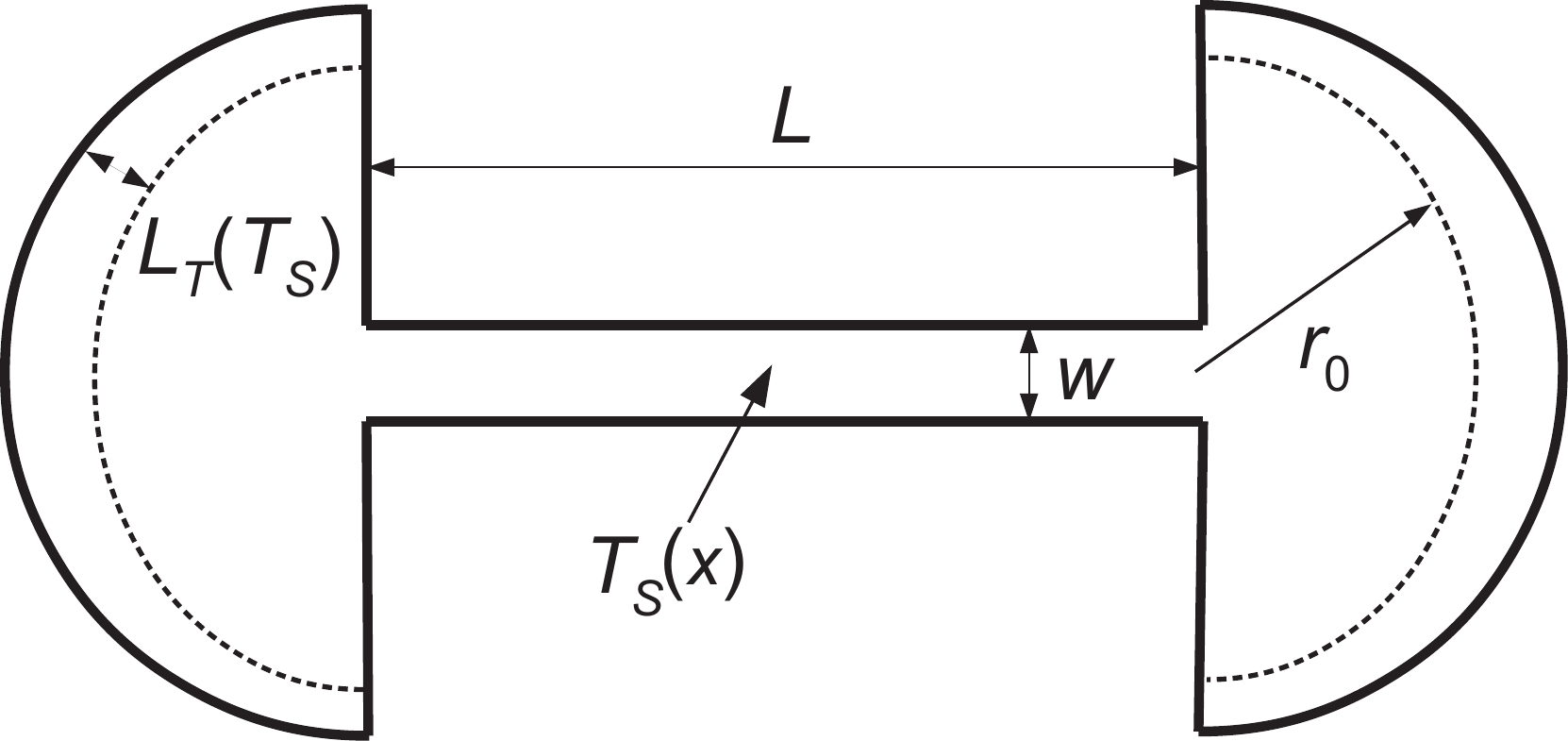}
	\caption{Model geometry of the graphene ribbon heater (size $L \times w$) which is connected to two wide,  half-infinite-plane graphene reservoirs at the ends. This geometry was employed for the analysis of the relaxation of $T_e$ outside the heater section. 
	%Current through the wire is denoted by  $I$ in the figure. 
	The radius $r_0$ determines the border of the region where electronic heat diffusion can be regarded as small.}
	\label{fig:wire}
\end{figure}

In order to evaluate the validity of the hot electron model, let us consider the behavior of electronic temperature at the ends of a wire coupled to two wide graphene leads as depicted in Fig. \ref{fig:wire}.

The temperature both in the ribbon and in the graphene leads can be found from the basic heat transport equation
\begin{eqnarray}
\frac{\pi^2 k_B^2 \sigma}{6e^2}\nabla^2 T_e^2 + \frac{j^2}{\sigma} - \Sigma_{sup} (T_e^3 - T_{ph}^3) = 0,
\label{lead}
\end{eqnarray}
where $\sigma$ is the conductivity per square, $j$ is the current per unit width ($j=I/w$) and the electron-phonon heat flow per unit area equals $\Sigma_{sup} = 7.5\times 10^{-4} \frac{D^2 \tilde{n}_s}{k_F \ell}$ according to the experimental supercollision results \cite{Song2012,Betz2012}; here $D \simeq 70$ eV denotes the deformation potential (in eV) and $\tilde{n}_s$ is the charge carrier density (in units of $10^{12}$ cm$^{-2}$). Using the above numbers, we obtain approximately equal heat fluxes for electron-phonon coupling in the heater and the electronic heat flux at the heater ends. Thus, $T_e^{\textrm{max}}$ becomes lowered to $\sim 50$ K.

Outside the narrow section of the heater, 
$j = \frac{I}{\pi r}$ in the graphene lead at a distance $r$ from the end of the ribbon.
At large bias, electron-phonon coupling dominates and the temperature in the graphene near the heater ends equals to
\begin{eqnarray}
T_S(r) = \left( T_{ph}^3 + \frac{j(r)^2}{\sigma\Sigma_{sup}} \right)^{1/3}.
\end{eqnarray}
A characteristic length scale for the decay of the heat flow along graphene reservoir can be obtained by equating the local variation of the gradient term in Eq. \ref{lead} to the change in heat flux due to electron-phonon coupling. The characteristic thermal relaxation length
scale $L_T(T_S)$ becomes
\begin{equation}
L_T(T_S)=\sqrt{\frac{\pi^2 k_B^2\sigma}{6e^2\Sigma_{sup} T_S}}. 
\label{LTT}
\end{equation}
At $T_S=1$ K, we obtain $L_T=2.3$ $\mu$m using a carrier concentration of $n_s=4 \times 10^{11}$ cm$^{-2}$ which is close to our experiments at a back gate voltage 5 V. Since this length decreases with temperature, it is evident that $T_e$ decreases quickly in the leads and that most of the heating power is deposited to the substrate within and near the heater. 

The Kapitza resistance between SiO$_2$ and graphene will enhance $T$ of graphene phonons upto $\sim2$ K, when the electronic temperature is 10 K. This will change slightly the above relaxation length, but it is still the supercollision cooling
with coupling $\Sigma_{sup} (T_e^3 - T_{ph}^3)$ that governs thermal relaxation of the electrons. 

\subsection{Thermal flow from the heater to the substrate}
On the basis of the smallness of the thermal relaxation length in Eq. \ref{LTT}, we assume that most of the heater power is deposited to the substrate within the area of the graphene ribbon heater and its immediate neighborhood. 
%Phonon thermal conductance along graphene is assumed to be good so that the power from graphene phonons enters SiO$_2$ over the whole area of the graphene reservoirs to the heater. 
Using 2 $\mu$m$^2$ for the thermal contact area, $A_K=500$ Wm$^{-2}$K$^{-4}$ for Kapitza conductance ($P=\frac{1}{4} A_K(T_h^4 - T_s^4)$ with temperatures $T_h$ and $T_s$ on the opposite sides of the interface) \cite{Swartz1989}, we obtain a phonon temperature of $T_h \sim 9$~K in graphene at 40 mV heating voltage. This elevated phonon temperature will relax along the substrate via phononic thermal conductance.

\subsection{Thermal flow on the SGS thermometer} \label{SGSflow}
The SGS thermometers are formed by two-terminal, Al/Ti-contacted graphene parts in the center of large pieces of graphene. They are made in such a way that there is an uninterrupted path of non-superconducting graphene connecting the thermometer to a quantum dot. This guarantees smooth thermal flow along the thermometer area without any strong suppression in the heat conductivity due to proximity-induced superconducting gap (note that the graphene electrode is fully crossed by a superconducting lead at the further end of the thermometer, see Fig. 1 of the main paper). The purpose of these SGS devices is to track the temperature of the electronic graphene reservoirs that govern the distribution of incoming electrons/holes on the graphene quantum dots. The large area of the SGS devices guarantees that their temperature will well track the temperature of the substrate, independent of a possible heat input coming along the graphene ribbons from the dots. 

The operating range of our SGS thermometers is up to 0.8 K, which facilitates thermopower studies up to rather substantial thermal gradients. At a typical thermometer temperature of 0.5 K, we obtain a characteristic relaxation length of 3.3 $\mu$m using Eq. \ref{LTT}. This is larger than the extent of thermometer along the heat flow direction. Consequently, we will take a spatially independent temperature for the SGS thermometer. In order to determine how the SGS thermometer averages over the spatially dependent substrate temperature $T_s(x)$, we assume that the heat flow balance between SGS and the substrate remains zero. This yields the integral condition
 $\int(T_{SGS}^3-T^3_s(x))dx=0$. 
 At low temperature with small $dT^3_s(x)/dx$, one may obtain a simple estimate for $T_{SGS} \simeq \left[\frac{1}{2}(T_l^3+T_r^3)\right]^{1/3}$ for the  temperature recorded by the thermometer; here $T_l$ and $T_r$ denote the temperature at the left and right edges of the thermometer region, respectively.

\subsection{Thermal flow along the substrate}

\subsubsection{Local phonon temperature and heat diffusion} \label{limits} 
For application of the regular heat diffusion equation, small mean free path of phonons is essential. In clean silicon, the mean free path of phonons can be on the order of 100 microns, which would lead to problems in defining a proper substrate temperature across our CPS device having a size smaller than 10 $\mu$m. However, the interfacial scattering at the SiO$_2$/Si interface and the strong impurity doping of the Si$^{++}$ material decrease the mean free path of phonons substantially. Typically in case of thin layers and rough surfaces at an interface, the phonon mean free path is limited to a value of the order of layer thickness \cite{Casimir1938} (Casimir limit). Thus, in our case $\ell_{ph}\sim 300$ nm in the SiO$_2$ that is in thermal contact with the CPS device. Owing to the relatively small $\ell_{ph}$, we argue that the heat diffusion equation can be employed to deduce approximate temperature distribution along the substrate across the area of the graphene reservoirs of our Cooper pair splitter.

\subsubsection{COMSOL simulations for temperature difference between the dots}

\begin{figure}[h]
	\includegraphics[width=1\textwidth]{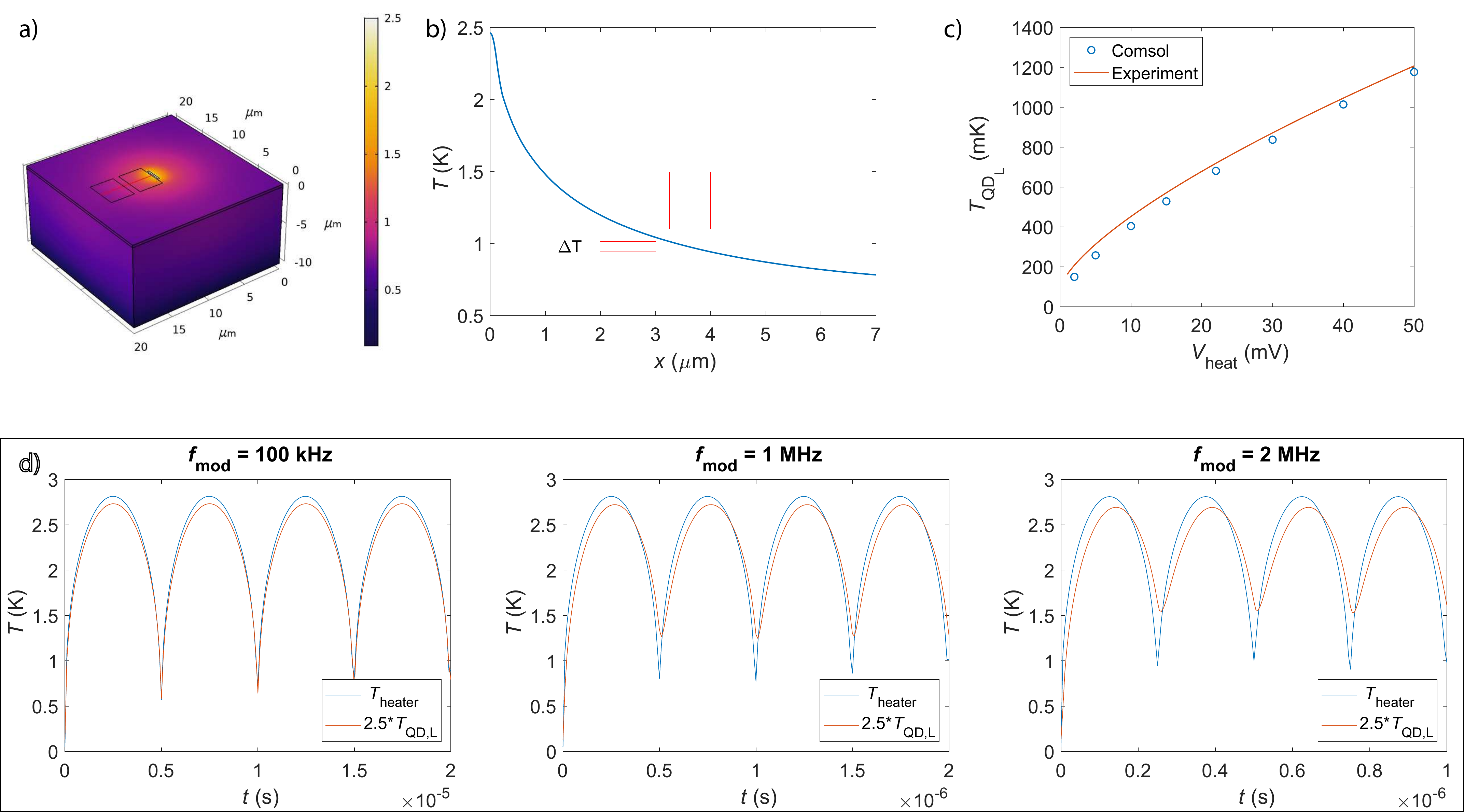}
	\caption{(a) 3D model used in the COMSOL simulations. The large block represents the 525 $\mu$m thick doped Si substrate, the layer on top of that is 300 nm of SiO$_2$, and the three rectangles on top of SiO$_2$ represent graphene pieces on top of the SiO$_2$ layer. Brighter (yellow) color indicates higher temperature.  (b) Simulated temperature at $V_h = 40$ mV depicted along the red line in (a) on the left, from the center of the heater through the quantum dots. The red markings indicate the locations of graphene reservoir regions (the edges of the area where the SGS thermometers are located) and the corresponding temperature drop $\Delta T \simeq 72$ mK. c) Temperature at the left quantum dot obtained from the Comsol model (blue dots), and experimental heater calibration of Eq. \ref{SGStemp} plotted as a function of the heater voltage. d) Temperature at the heater (blue) and $T_{QD,L}$ (red) presented as a function of time when the heater power is modulated sinusoidally at maximum heating power ($V_h = 40$ mV) using frequencies $f =$ 0.1 MHz, 1 MHz, and 2 MHz indicated on top of the figures.}
	\label{fig:Comsol_simulations}
\end{figure}

The heat diffusion equation was solved using COMSOL multiphysics. The geometrical model used in the simulations (see Fig. \ref{fig:Comsol_simulations}) is a simplified version of the real device presented in Fig. 1 of the main text. 
 %\ref{fig:SampleIllustration}. 
 Our simulations using exact sample dimensions focused on the temperature profile on the surface along the line drawn through the heater and both quantum dots. 
 Best available data on materials parameters were employed in calculating thermal gradients along the substrate. The graphene heater power was inputted only over the actual heater size, which slightly increased the maximum temperature of the SiO$_2$ but this was deemed irrelevant further away from the heater which was the main region of interest. 
 
 In order to achieve most realistic temperature profile, we employed temperature dependent heat conductivities both for the phonon thermal conductance as well as for the metallic heat diffusion. The mean free path of phonons influences strongly the thermal conductivity (see Sect. \ref{limits}). We employed 
$\kappa_{SiO_2} = 0.01 \times T^2$ W/(mK$^3$) \cite{Zeller1971},
$\kappa_{gr} =0.02 \times T^2$ W/(mK$^3$) \cite{PoP2012,Seol2010}, and 
 $\kappa_{Si^{++}} =0.01 (T/K)^2$ W/mK$^3$ \cite{Slack1964}  for thermal conductivities of  SiO$_2$, graphene, and strongly doped silicon, respectively. For dynamical calculations, we also specified heat capacities $C_{SiO_2} =3 \times 10^{-3} \times T^2$ J/(kgK$^3$), $C_{gr} =1 \times T$ J/(kgK$^2$), and $C_{Si^{++}} =10^{-3}$ J/(kgK$^3$), respectively.
 
  %Typical thermal transport parameters around $T=1$ K (heat capacitance and thermal conductivity) for the materials involved (graphene, SiO$_2$, and Si) were put into the simulation.

Fig. \ref{fig:Comsol_simulations} displays simulation results for the spatially dependent temperature $T(x)$ on the surface of the SiO$_2$ layer as a function of the distance  $x$ from the heater along the central symmetry axis intersecting the two quantum dots and the center of the graphene ribbon heater. The simulated trace at $V_h = 40$ mV yields  $\Delta T = 72$ mK for the temperature difference between the two red lines marking the edges of the SGS graphene thermometers. The flatter regions before and after the steep section between the red lines is due to the large phononic conductance of graphene. Note that the regions would be even flatter if the electron-phonon coupling and the electronic heat conductance $\kappa_e$ in graphene would have been included. This omission of $\kappa_e$ also renders the modification of electrical thermal conductivity by the induced superconducting gap irrelevant for the simulation. 
%Using the heat balance equation of Sect. \ref{SGSflow} for both SGS thermometers, we find $\Delta T \simeq 200$ mK, which is quite well in line with the difference indicated by the SGS thermometers. 

The COMSOL simulation was also used to estimate the thermal time constant of the system. This was done by looking at the phase difference between the heater temperature $T_{h}$ and $T_{QD_L}$ as the heater modulation frequency was increased. The results are presented in Fig. \ref{fig:Comsol_simulations}d at frequencies 0.1, 1, and 2 MHz. The simulation demonstrates that at 100 kHz the phase difference is still negligible, while at higher frequencies it starts to grow. Thus, we can conclude that the time constant $\tau < 0.01$ ms and steady state approximations can safely be applied to our measurements at $f<10$ Hz. 

%\begin{figure}[h]
%	\includegraphics[width=1\textwidth]{Fig_S4.pdf}
	%\includegraphics[width=0.32\textwidth]{pics/frequencyResponse_100kHz.pdf}
	%\includegraphics[width=0.32\textwidth]{pics/frequencyResponse_200kHz.pdf}
%	\caption{Temperature at the heater (blue) and $T_{QD,L}$ (red) presented as a function of time when the heater power is modulated sinusoidally at maximum heating power ($V_h = 40$ mV) using frequencies $f =$ 10 kHz, 100 kHz, and 200 kHz indicated on top of the figures.}
%	\label{fig:Comsol_simulationsTimeconstant}
%\end{figure}

%As seen in the figure, the chain of heat transport from the electron system of the graphene heater to the quantum dots is quite complicated, and consequently analytical modeling of the system turned out very uncertain and challenging.

%
%\begin{minipage}[h]{0.5\textwidth}
%In order to get a rough estimate of the temperatures induced by the heater, we consider semi-1D heat transport facilitated by Weidemann-Franz heat diffusion:
%\begin{align}
%\nabla(\kappa\nabla T_e) &= P_{Joule} \\
%\nabla(T_e \nabla T_e) &= \frac{V_{heat}^2/R}{\sigma L}
%\label{eq:heatEq}
%\end{align}
%where $\kappa = \sigma LT_e$ was invoked to get the second line while assuming temperature independent electrical conductivity $\sigma$.
%
%One can solve the simplified Eq. \ref{eq:heatEq} to get:
%\begin{equation}
%T_{e,max} = \sqrt{\frac{V_{heat}^2/R}{4\sigma L}}.
%\end{equation}
%This is plotted on the right for the heater range used in the experiments.
%\end{minipage}
%\begin{minipage}[h]{0.5\textwidth}
%\includegraphics[width=\textwidth]{pics/analyticalTe.pdf}
%\end{minipage}

\clearpage
\setcounter{subsection}{0}
\setcounter{subsubsection}{0}
\section*{Note 2: Calibration of heater voltage vs. SGS junction temperature}

Once calibrated, the heater voltage could be used for specifying the temperature difference across our Cooper pair splitter. This was useful, in particular because of the poor resolution of the $SGS_R$ thermometer, which imposed long measurement times for good accuracy. The poor resolution of $SGS_R$ was due to lack of measurement wires in the cryostat, which forced us to connect the $SGS_R$ thermometer in a 2-wire configuration.

\begin{figure}[h!]
\includegraphics[width=\textwidth]{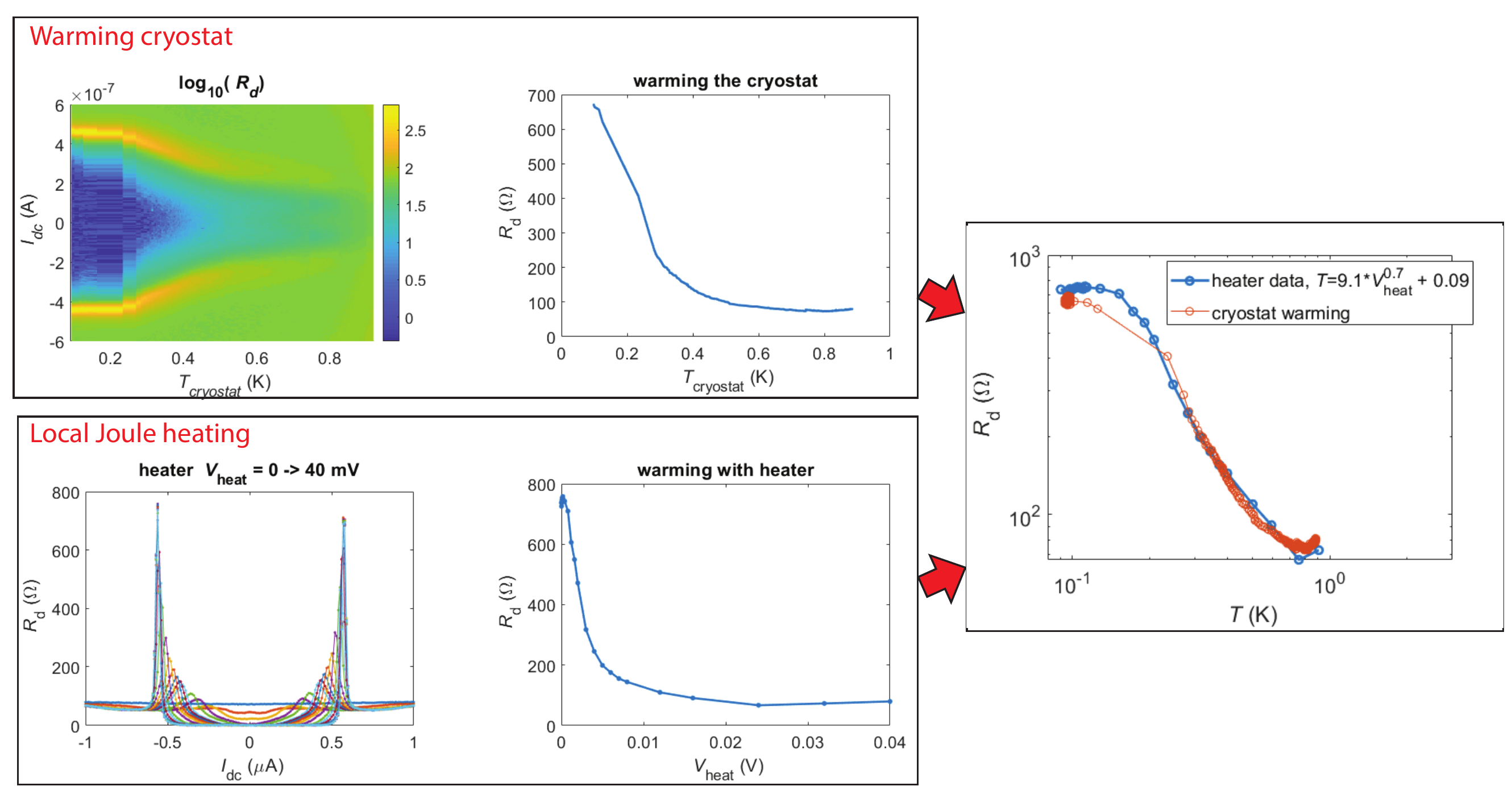}
\caption{ Measurement data sets employed for obtaining a relation between the heater voltage $V_{h}$ and the SGS junction temperature $T_{SGS_L}$. Upper left frame: (left) Logarithm of the differential resistance of the $SGS_L$ junction displayed on the cryostat temperature vs. dc bias current plane. (right) Peak resistance value of $R_{d}$ vs. equilibrium temperature $T_{cryostat}$ while warming the cryostat. Lower left frame: (left) Differential resistance of the $SGS_L$ junction vs. dc bias current at different heating powers applied to the galvanically-separated graphene ribbon heater ($V_h = 0 \dots 40$ mV). (right) Peak resistance value $R_d^{peak}=\max{R_{d}(I_{DC})}$ vs. heating voltage. Right frame: Data for the peak value $R_d^{peak}$ scaled to match the maximum sub-gap value of $R_{d}(T)$ using Eq. \ref{SGStemp}. }
\label{fig:heaterCalibrationScheme}
\end{figure}

The temperature of $SGS_L$ $w.r.t.$ the heater voltage was calibrated by measuring the IV characteristics of the $SGS_L$ as a function of the heater voltage $V_{h}$ and the equilibrium cryostat temperature $T_{cryostat}$. The obtained differential resistance $R_d$ characteristics are presented in the left frames of Fig. \ref{fig:heaterCalibrationScheme}, in which the decrease of the superconducting gap with increasing  $V_{h}$ and $T_{cryostat}$ can be seen clearly. At the edge of the gap, we traced the maximum differential resistance, $R_d^{peak}=\max{R_{d}(I_{DC})}$, both as a function of $V_{h}$ and $T_{cryo}$. The observed traces can be made to overlap each other by adjusting the x-scale, i.e. by mapping the voltage scale to a corresponding temperature scale. The obtained scaling relation for $SGS_L$ is given by:
\begin{equation}
T_L = (9.1 \cdot V_{h}^{0.7} + 0.09) \textrm{ K},
\label{SGStemp}
\end{equation}
where $V_{h}$ is in volts. Note that there is an "offset" of 90 mK that we interpret as the effective electron temperature in graphene at the cryostat base temperature when no heating is applied. For $SGS_R$, we obtain the relation $T_R = (8.4\cdot V_{h}^{0.7} + 0.09) \textrm{ K}$ .

Note that the exponent of the heater voltage, 0.7, is approximately consistent with $T_L^3 \propto  V_{h}^2$ (and $T_R^3 \propto  V_{h}^2$) that would in turn be consistent with the cooling $P \propto T^3$ being dominated by acoustic phonons, for which the thermal conductivity $\kappa_{ph} \propto T^3$.\\

\clearpage
\section*{Note 3: Additional data on bias-induced thermopower}

 The thermoelectric current measured in the main paper was purely driven by temperature gradient, without any bias applied on the Cooper pair splitter. We also performed experiments where the thermoelectric response was investigated without a separate heater. Bias current in the graphene ribbon connecting the QDs to the graphene reservoirs leads to heating which creates basically local thermopower phenomena with symmetry properties close to the non-local phenomena. Because of this, the central figures in the main paper (Figs. 3 and 4) contain only data with plain thermal drive.   
 
\begin{figure}[ht]
\begin{center}
	\includegraphics[width=14cm]{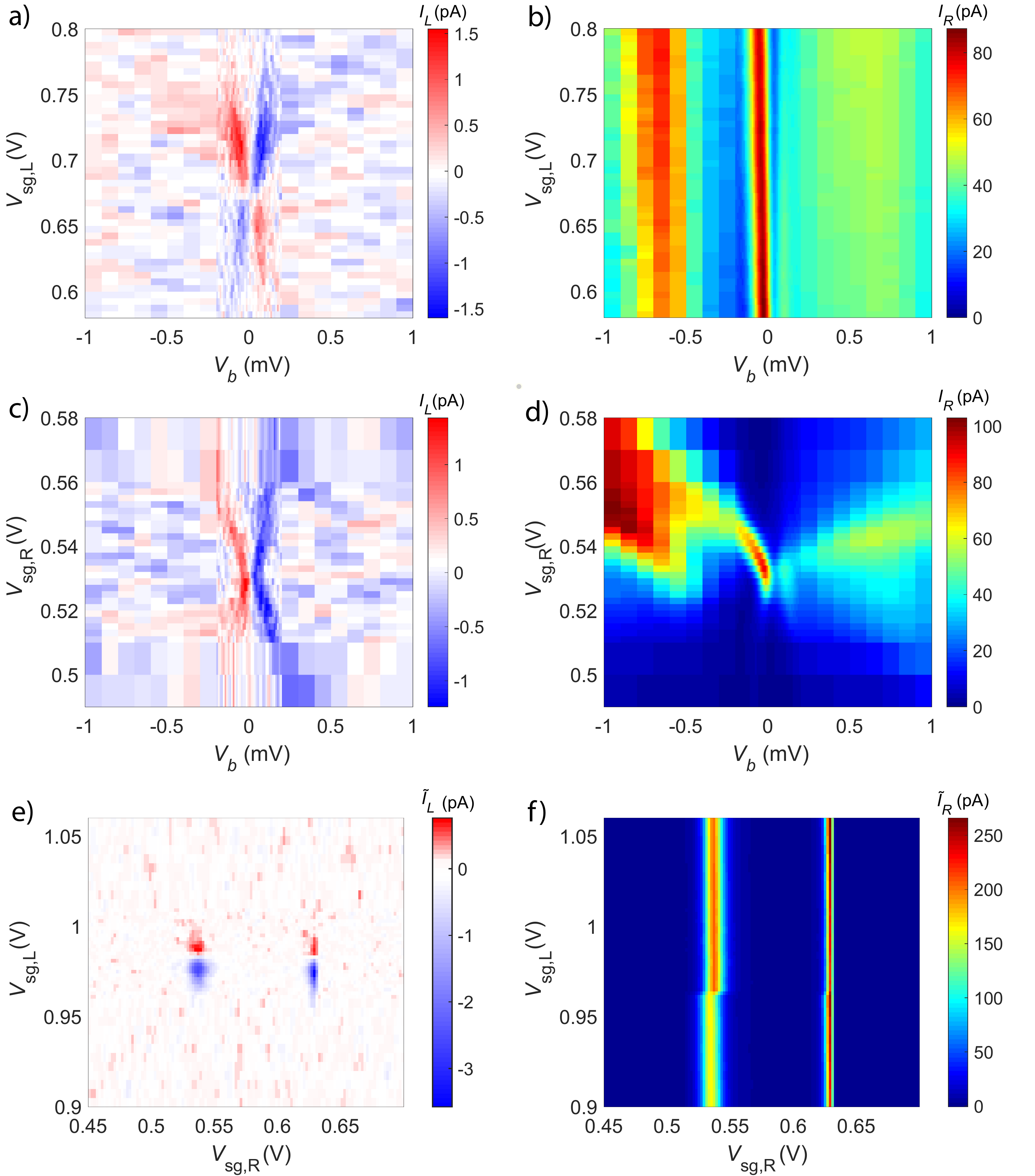}
	\caption{ Splitting at finite bias: competition of CPS, EC and local thermoelectric effects. (a) Ac current measured on QD$_L$ as a function $V_{sg,L}$ while sweeping the dc bias $V_b$ on QD$_R$ at $V_{sg,R}=0.535$ V (b) Ac current of QD$_R$ as a function of $V_{sg,L}$ while sweeping dc bias on QD$_R$ at $V_{sg,R}=0.535$ V. (c) Ac current of QD$_L$ as a function of $V_{sg,R}$ while sweeping dc bias on QD$_R$ at $V_{sg,L}=0.72$ V. (d) Ac current of QD$_R$ as a function of $V_{sg,R}$ while sweeping dc bias on QD$_R$ at $V_{sg,L}=0.72$ V. In all of these measurements a-d, QD$_R$ was also biased using 20 $\mu$V$_{rms}$ at 2.1 Hz for ac measurements. (e) Apparent non-local thermoelectric effects due to variation in Joule heating across the resonance. Current $\tilde{I}_L$ (at $2\omega$) of QD$_L$ measured over $V_{sg,L}$ and $V_{sg,R}$ plane using an ac bias of 50 $\mu$V on QD$_R$. (f) The corresponding current $I_R$ of QD$_R$ measured at $\omega$ over $V_{sg,L}$ and $V_{sg,R}$ plane. Notice the resemblance of frame (e) with the non-local thermoelectric current patterns in Fig. 3 of the main paper. Here, the large ac excitation leads to gate voltage dependent heating in QD$_R$, which leads to variation of $T$ with the same symmetry properties as the non-local thermoelectric effect with constant thermal gradient.	
	}
	\label{fig:condHEAT}
\end{center}
\end{figure}

 Fig. \ref{fig:condHEAT} shows results on thermoelectric current in a Cooper pair splitter driven by a voltage bias  on one of the quantum dots. In this case, thermoelectric effects and the bias-induced CPS and EC will mix together in the behavior. Fig. \ref{fig:condHEAT}a and \ref{fig:condHEAT}b display the currents of the two dots $I_{L}$ and $I_{R}$ measured at $\omega$ as function of the dc bias $V_{b}$ and the side gate voltage $V_{sg,L}$ on QD$_L$; here dc and ac voltage bias is applied only on QD$_R$ with the middle superconcucting lead grounded. The current $I_{L}$ shows different sign with positive and negative dc bias $V_{b}$ on QD$_R$. In addition, $I_{L}$  changes its sign when $V_{sg,L}$ crosses $V_{sg,L}=0.72$ V, at which the energy level of QD$_L$ matches the Fermi energy. Ac current $I_{R}$, however, is insensitive to the $V_{sg,L}$. 

Figs. \ref{fig:condHEAT}c and \ref{fig:condHEAT}d display $I_{L}$ and $I_{R}$  as functions of dc bias $V_{b}$ on QD$_R$ and the side gate voltage $V_{sg,R}$. Here $I_{L}$ changes its sign, when $V_{b}$ crosses zero, but it retains its sign when $V_{sg,R}$ crosses $V_{sg,R}=0.535$ V, at which the energy level of QD$_R$ matches the Fermi level. Note that the observed variation of $I_{L}$  as function of $V_{sg,L}$, $V_{sg,R}$ and $V_{b}$ can be explained by considering heating due to $V_b$ and local thermoelectric currents induced by the ensuing thermal gradients between QD$_L$ and the superconductor. Unfortunately, there are no means to single out the non-local thermoelectric effects in these sets of data.

If QD$_R$ is ac biased at $\omega$ without any dc, we obtain Fig. \ref{fig:condHEAT}e for the current $\tilde{I_L}$ in QD$_L$ at $2\omega$. The thermoelectric current in QD$_L$ changes its sign while the energy level tuned by $V_{sg,L}$ crosses the Fermi energy. On the contrary, $\tilde{I_L}$ keeps its sign as the energy level is tuned by $V_{sg,R}$ across the Fermi surface. These results are consistent with those in Figs. \ref{fig:condHEAT}a-d. 

Although the observed data in Fig. \ref{fig:condHEAT} are similar to those in Figs. 3 and 4 in the main paper, and apparently non-local effects according to this similarity in symmetry and asymmetry, they can be explained using bias-induced heating and local thermoelectric effects induced by thermal gradients. In Figs. \ref{fig:condHEAT}a and \ref{fig:condHEAT}c, the dc and ac bias applied to QD$_R$ causes heating in the middle Al lead. When the dc bias is positive, the heating in the Al lead will be in the same phase with the ac bias of QD$_R$: i.e. the heating in Al follows the maximum and minimum of the ac bias on QD$_R$. When the dc bias is negative, however, the maximum heating will coincide with minimum of the ac, and the heating in the middle Al will be phase shifted by $\pi$ from the ac bias on QD$_R$. Hence, in Figs. \ref{fig:condHEAT}a and \ref{fig:condHEAT}c, the thermoelectric current $I_L$ has an opposite sign at positive or negative dc bias. In Fig. \ref{fig:condHEAT}a,  $I_L$ changes its sign due to the same reason as in Fig. 2a (main text) when $V_{sg,L}$ crosses $V_{sg,L}=0.675$ V. Because the heating in the middle Al lead is equal at left and right fo the conductance peak of QD$_R$, $I_L$ keeps the same sign while energy level of QD$_R$ crosses Fermi level in Fig. \ref{fig:condHEAT}c.

%\begin{figure}[H]
%	\includegraphics[width=16cm]{Fig_S7.pdf}
%	\caption{Apparent non-local thermoelectric effects due to variation in Joule heating across the resonance. (a) Current $\tilde{I}_1$ (at $2\omega$) of QD1 measured over $V_{sg1}$ and $V_{sg2}$ plane using an ac bias of 50 $\mu$V on QD2. (b) The corresponding current $I_2$ of QD2 measured at $\omega$ over $V_{sg1}$ and $V_{sg2}$ plane. Notice the resemblance of frame (a) with the non-local thermoelectric current patterns in Fig. 3 of the main paper. Here, the large ac excitation leads to gate voltage dependent heating in QD2, which leads to variation of $T$ with the same symmetry properties as the non-local thermoelectric effect with constant thermal gradient.}
%	\label{fig:condarms}
%\end{figure}

Even when using the $2\omega$ method, driving QD$_R$ at $\omega$, measuring $I_L$ at $2\omega$, the same heating argument can be employed to explain the observed bahavior in Fig.  \ref{fig:condHEAT}e. In the absence of bias voltage, the heating is now at $2\omega$, at which the signal then appears. These data have strong resemblance with the result in Fig. 4d in the main paper. The current $I_R$ at $\omega$ is depicted in Fig.  \ref{fig:condHEAT}f. 

% \section{Quantum dot model for non-local thermoelectric currents}

\clearpage
\setcounter{subsection}{0}
\setcounter{subsubsection}{0}
\section*{Note 4: Theoretical modeling: Coherent transport in N-Dot-S-Dot-N system}

\begin{figure}[!ht]
	\includegraphics[width=9cm]{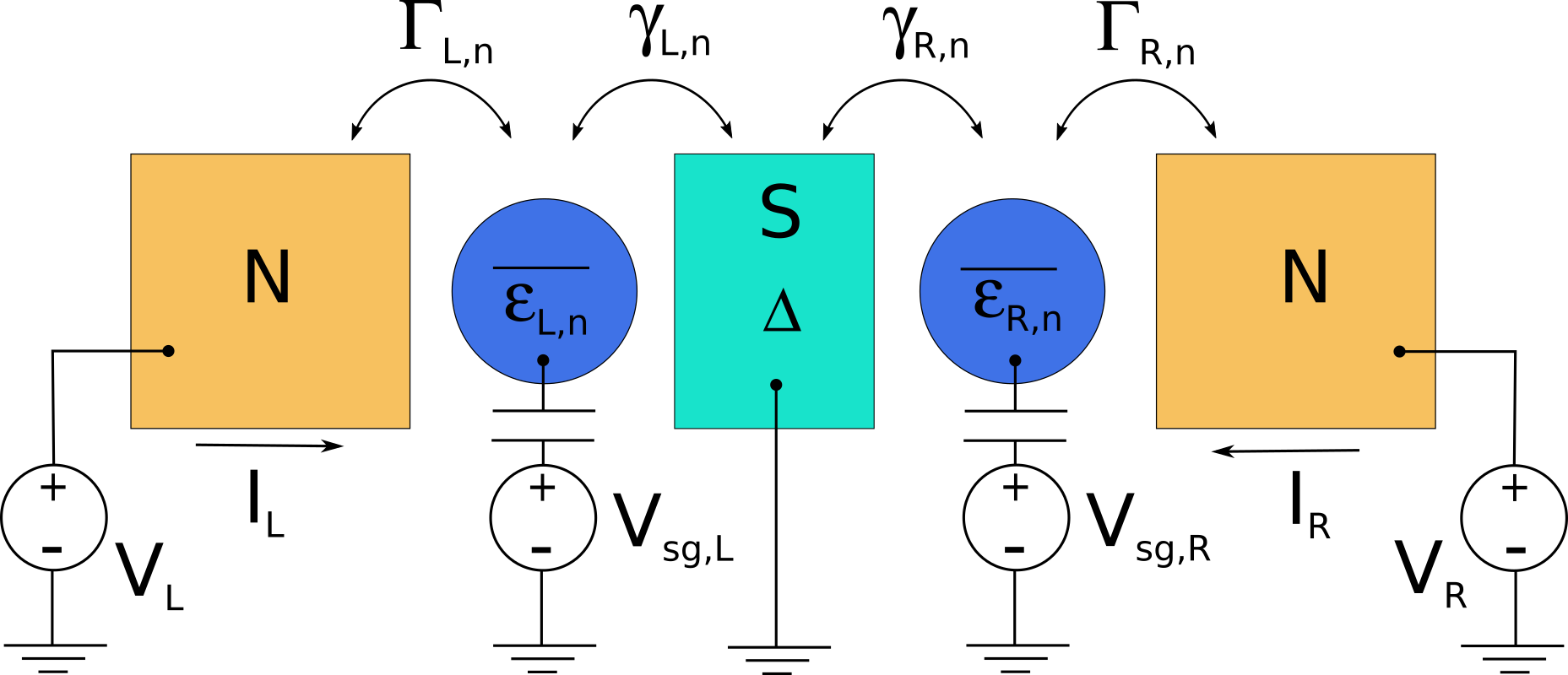}
	\caption{Schematics of an N-dot-S-dot-N system. Two quantum dots with the energy levels $\varepsilon_{L,n}$ and $\varepsilon_{R,n}$
     are coupled to the common superconducting lead with the gap $\Delta$ and the temperature $T_S$, and 
     to the two different normal bulk leads with the temperatures $T_L$ (left lead) and $T_R$ (right lead). The tunneling rates between the dots
     and the superconductor are $\gamma_{L,n}$ and $\gamma_{R,n}$, while the tunneling rates between the dots and the normal leads are $\Gamma_{L,n}$ 
     and $\Gamma_{R,n}$. Bias volatges $V_L$ and $V_R$ can be applied to the outer leads, and the side gate potentials $V_{sg,L}$ and $V_{sg,R}$
     allow one to shift the energy leves $\varepsilon_{L,n}$ and $\varepsilon_{R,n}$. The potential of the superconductor is supposed to be zero.
     The electric currents $I_L$ and $I_R$ flow through the quantum dots into the superconductor.} 
	\label{NDSDN}
\end{figure}

In this note we outline the theory of thermal transport in the system  
consisting of two quantum dots coupled to two separate normal leads
and one common superconducting lead. The schematics of it is shown in Fig.\,\ref{NDSDN}.
Here, we describe the coherent transport regime, which rather well describes the main findings of our
experiment. 
The additional sign changes in
the local thermal current at low heating power which are not captured by the present description are explained within the incoherent model discussed in Note 5. 

We assume that the conductances of the dots are high enough so that 
the dots enter the Fabry-P\'erot regime, and one can ignore Coulomb blockade.
We split the energy levels in both dots into two groups: well coupled to the leads levels $\varepsilon_{j,n}$ 
(here the index $j=L,R$ indicates the left and right dot, respectively, and $n$ enumerates the energy levels within one dot), and poorly coupled to the leads levels $\varepsilon'_{j,n}$.   
The hopping rates of electrons between the levels $\varepsilon_{j,n}$ and the normal leads,  which in our experiment are big graphene flakes,
are denoted by $\Gamma_{j,n}$, while the hopping between the dots and the superconducting lead is set by rates $\gamma_{j,n}$.
The energy levels can be tuned by side gate potentials $V_{sg,j}$ as 
\begin{eqnarray}
\varepsilon_{j,n}(V_{sg,j}) = a_j (V_{sg,j}-V_{\max,j,n}),\;\;\; \varepsilon'_{j,n}(V_{gj}) = a_j (V_{sg,j}-V'_{j,n}).
\end{eqnarray}
Here the constants $a_j$ are determined by the ratios of the gate and the dot capacitances, and $V_{\max,j,n}$ are the
gate voltages at which the conductance of the corresponding dot exhibits a peak and reaches a local maximum, and $V'_{j,n}$ 
are the gate voltages at which the energies of the "dark" states $\varepsilon'_{j,n}$ become equal to zero.
From the experiment we approximately find
\begin{eqnarray}
a_L\approx 0.003\;\; \frac{{\rm eV}}{{\rm V}},\;\;\, a_R\approx 0.01\;\; \frac{{\rm eV}}{{\rm V}}.
\label{alpha}
\end{eqnarray}
Bias voltages $V_j$ are applied to the metallic leads, while the potential of the superconductor equals zero.

\subsection{Local electric current in the normal state}

In this section, we recall some well known results and consider fully normal system assuming $\Delta=0$.
Since we neglect the Coulomb blockade, transport of charge through the quantum dots is well described by the Landauer formula,
\begin{eqnarray}
I_j^{\rm loc} = \frac{2e}{h}\int dE\,\tau_j(E,V_{sg,j})[f_j(E-eV_j,T_j) - f_S(E,T_S)],
\label{Landauer}
\end{eqnarray}
where the superscript "loc" indicates the local character of the current (i.e. $I_j^{\rm loc}$ can be evaluated assuming that the other quantum dot does not exist),
$h$ is Planck's constant, 
$\tau_j(E,V_{sg,j})$ is the transmission probability of the dot,
\begin{eqnarray}
f_j(E-eV_j,T_j) = \frac{1}{1+e^{(E-eV_j)/k_BT_j}},\;\;\; f_S(E,T_S) = \frac{1}{1+e^{E/k_BT_S}}
\end{eqnarray}
are the electron distribution functions in the bulk normal lead $j$ and in the superconductor, $V_j$ is the electric potential applied to the lead $j$, 
$T_j$ is the temperature of the lead $j$, and $T_S$ is the temperature of the central electrode.  
The factor $2$ in front of Eq. (\ref{Landauer}) accounts for the spin degeneracy.
We use two models for the transmission probability: (i) simple resonant tunneling model, in which 
$\tau_j(E,V_{sg,j})$ is given by a sum of Lorentzian peaks, each coming from one of the energy levels,
\begin{eqnarray}
\tau_j(E,V_{sg,j}) = \sum_{n} \frac{\gamma_{j,n}\Gamma_{j,n}}{\left(E-\varepsilon_{j,n}(V_{sg,j})\right)^2 + \frac{(\gamma_{j,n}+\Gamma_{j,n})^2}{4}},
\label{tau_res}
\end{eqnarray}
and (ii) a model with Fano resonances, in which
\begin{eqnarray}
\tau_j(E,V_{sg,j}) = \sum_{n} \frac{\gamma_{j,n}\Gamma_{j,n}}{\left(E-\varepsilon_{j,n}(V_{sg,j})-\frac{|t_{j,n}|^2}{E-\varepsilon'_{j,n}(V_{sg,j})}\right)^2 + \frac{(\gamma_{j,n}+\Gamma_{j,n})^2}{4}}.
\label{tau_Fano}
\end{eqnarray}
Here $t_{j,n}$ is the hopping amplitude between the energy level $\varepsilon_{j,n}$, which is coupled to the leads, and to one of the closely lying
uncoupled (dark) levels with the energy $\varepsilon'_{j,n}$. Of course, for some  peaks one can have $t_{j,n}=0$, and in this case the transmission has the same 
Lorentzian shape as in Eq. (\ref{tau_res}). 
 
According to Eq. (\ref{Landauer}), the zero-bias dimensionless conductance of the dot in the zero temperature limit has the form
\begin{eqnarray}
g_j(V_{sg,j}) = \frac{h}{e^2}\frac{dI_j}{dV_j}\bigg|_{V_j=0} = 2\tau(0,V_{sg,j}).
\end{eqnarray}
In the limit $k_BT_j,k_BT_S \ll \Gamma_{j,n}+\gamma_{j,n}$ and for $V_j=0$, Eq. (\ref{Landauer}) reproduces the Mott formula for thermal currents
\begin{eqnarray}
I_j = \frac{\pi^2}{3}\frac{ek_B^2}{h}\frac{\partial\tau(0,V_{sg,j})}{\partial E}(T_j^2-T_S^2).
\end{eqnarray}
The derivatives of the transmission probabilities over the energy $E$ can be converted into the derivatives over the gate potentials
if one uses Eqs. (\ref{tau_res},\ref{tau_Fano}) in combination with the gate efficiency relations in Eq. (\ref{alpha}). In this way, one finds
\begin{eqnarray}
I_j = -\frac{\pi^2}{6}\frac{ek_B^2}{a_j h}\frac{\partial g(V_{sg,j})}{\partial V_{sg,j}}(T_j^2-T_S^2).
\label{Ij_app}
\end{eqnarray}
The corresponding thermopower, or the Seebeck coefficient, is given by
\begin{eqnarray}
\alpha_j = \lim_{T_j\to T_S} \frac{\Delta V_j}{T_j-T_S}\bigg|_{I_j=0} = \lim_{T_j\to T_S} \frac{h I_j}{e^2 g_j(V_{sg,j})(T_j-T_S)}\bigg|_{V_j=0}
= \frac{\pi^2}{3} \frac{k_B^2T_S}{ea_j} \frac{1}{g_j(V_{sg,j})}\frac{\partial g_j(V_{sg,j})}{\partial V_{sg,j}}.
\label{Seebeck}
\end{eqnarray}
Eqs. (\ref{Ij_app}, \ref{Seebeck}) allow us to estimate the expected values of thermoelectric currents and the thermopower once
the dependence of the zero bias conductance on the gate voltage has been measured. Eq. (\ref{Ij_app}) also determines the sign of the thermoelectric current and
the values of the gate voltage at which it equals to zero.

\subsection{Local electric current in the superconducting state}

In this section, we assume that  the central lead of the system  shown in Fig. \ref{NDSDN} is superconducting.
In this case, the local transport properties of the individual dots are described by the theory of Andreev reflection at NS interfaces \cite{Blonder1982,Beenakker1992},
which needs to be generalized to the case of energy dependent transmission probability. 
For the resonant tunneling model, such generalization has been presented, for example, in Ref. \onlinecite{Sun1999}.
The result of Ref. \onlinecite{Sun1999} can be transformed to the form
\begin{eqnarray}
I_j^{\rm loc}(E,V_{sg,j}) &=& \frac{e}{\pi\hbar}\int_{|E|<\Delta} dE \frac{\Delta^2[f_j(E-eV_j)-f_j(E+eV_j)]}
{(\Delta^2-E^2)\left[ \left(\frac{2}{\tau_j\left(\tilde E_j,V_{sg,j}\right)}-1\right)\left(\frac{2}{\tau_j\left(-\tilde E_j,V_{sg,j}\right)}-1\right) 
-\frac{4\tilde E_j^2}{\Gamma_j^2(V_{sg,j})}\right]+E^2}
\nonumber\\ &&
+\, \frac{2e}{\pi\hbar}\int_{|E|>\Delta}dE\frac{\nu_S(E)\left(\frac{2}{\tau(-E,V_{sg,j})}-1+\nu_S(E)\right)[f_j(E-eV_j)-f_S(E)]}
{\left(\frac{2}{\tau(-E,V_{sg,j})}-1+\nu_S(E)\right)\left(\frac{2}{\tau(E,V_{sg,j})}-1+\nu_S(E)\right)+\frac{4\Delta^2}{\Gamma_j^2(V_{sg,j})}\nu_S^2(E)},
\label{Ij}
\end{eqnarray}
which is convenient for the analysis of the experimental data.
Here we have introduced the re-normalized energy  $\tilde E$ and the density of states in the superconductor $\nu_S(E)$,   
\begin{eqnarray}
\tilde E_j = E\sqrt{1+\frac{\gamma_j(V_{sg,j})}{\sqrt{\Delta^2-E^2}}},\;\;\; \nu_S(E)=\frac{|E|}{\sqrt{E^2-\Delta^2}}.
\end{eqnarray}  
We have also replaced the hopping rates of the individual levels $\Gamma_{j,n}$ and $\gamma_{j,n}$ by the gate-voltage-dependent
expressions $\Gamma_j(V_{sg,j})$ and $\gamma_j(V_{sg,j})$. 
In fact, Eq. (\ref{Ij}) provides a full description of the local transport through the quantum dots, and it can be used beyond
the resonant tunneling model. For example, 
Eq. (\ref{Ij})  accounts fully for the contribution of Andreev bound states, which are formed in the dots due to their coupling
to a superconducting lead, and which have been observed  in InAs \cite{Deacon2015} and in carbon nanotube quantum dots \cite{Gramich2017}.
If the transmissions $\tau_j$ do not depend on the energy and $\Gamma_j\gg\Delta$,  
Eq. (\ref{Ij}) reduces to well known result by Blonder Tinkham and Klapwijk \cite{Blonder1982}.
In the limit $\Delta=0$, Eq. (\ref{Ij}) takes the Landauer form (\ref{Landauer}). 
Finally, according to Eq. (\ref{Ij}) in the low temperature limit,  $k_BT_j,k_BT_S \ll \Delta, \Gamma_{j,n}+\gamma_{j,n}$, 
the zero bias conductance of the dot $j$ acquires the form 
\begin{eqnarray}
g_j(V_{sg,j})= \frac{h}{e^2}\frac{\partial I_j}{\partial V_j}\bigg|_{V_j=0} = \frac{4\tau^2_j(0,V_{sg,j})}{\left(2-\tau_j(0,V_{sg,j})\right)^2},
\label{Andreev}
\end{eqnarray}
which, as expected, coincides with the prediction of the theory of Andreev reflection at low energies\cite{Beenakker1992}. 

For an asymmetric quantum dot with low transmission, $\tau_j(E)\ll 1$,
and for $eV,k_BT_j,k_BT_S,\Gamma_j,\gamma_j\ll \Delta$, Eq. (\ref{Ij}) reduces to a simple expression
\begin{eqnarray}
I_j^{\rm loc}(E,V_{sg,j}) = \frac{e}{2\pi\hbar}\int dE \, \tau_j(E,V_{sg,j})\tau_j(-E,V_{sg,j}) \, [f_j(E-eV_j)-f_j(E+eV_j)],
\label{subgap}
\end{eqnarray}
which contains the product of the transmission probabilities for an incoming electron, $\tau_j(E,V_{sg,j})$, and for a reflected hole,
$\tau_j(-E,V_{sg,j})$.

\begin{figure}[!ht]
		\includegraphics[width=14cm]{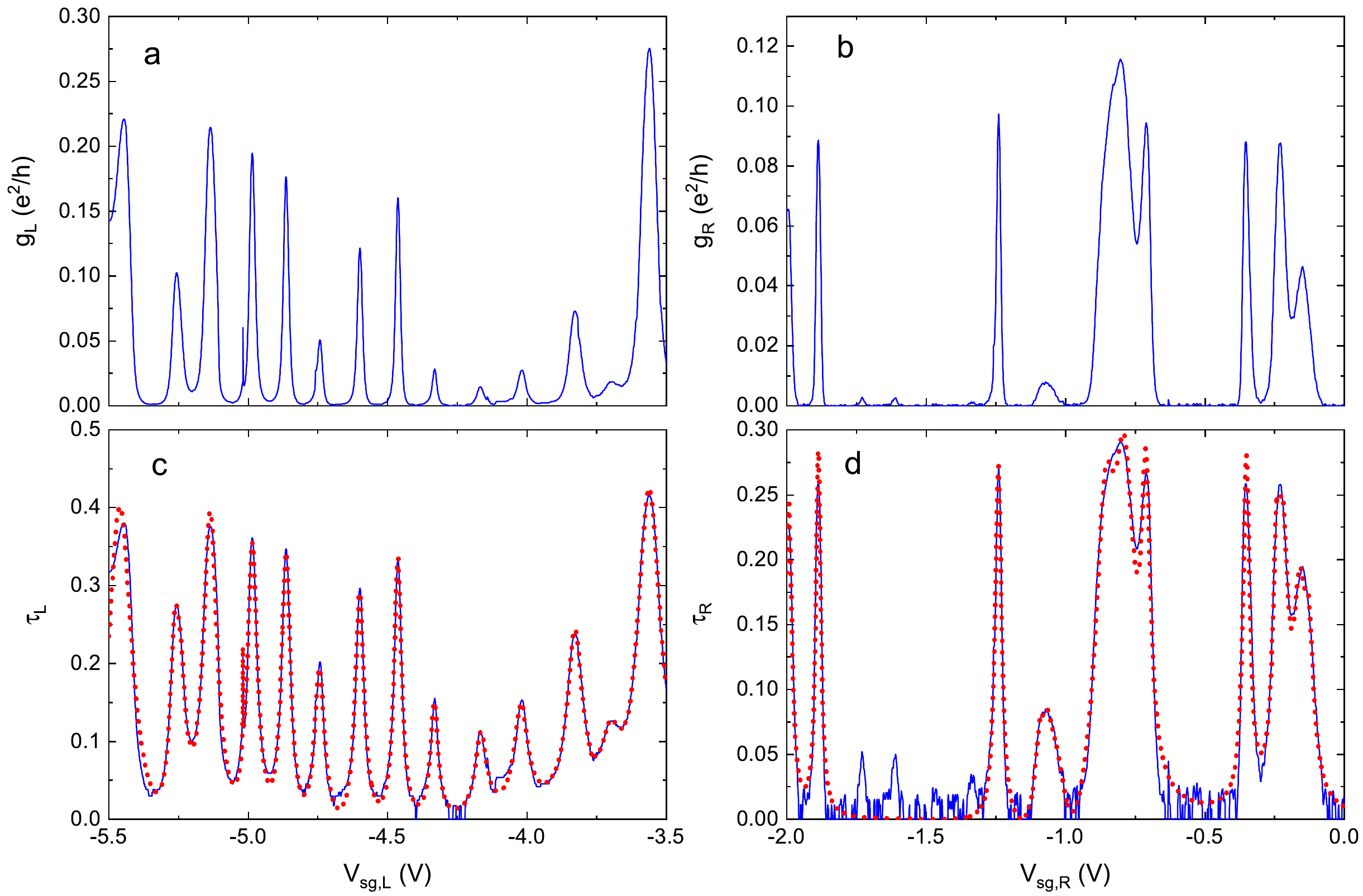}
	\caption{(a) and (b) --- measured zero bias conductance of the left and right quantum dots. 
            (c) and (d) --- normal state transmission probabilities $\tau_L(V_{sg,L})$ and $\tau_R(V_{sg,R})$
		obtained by applying the transformation of Eq. (\ref{Tj}) to the experimental conductance curves  depicted in panels (a) and (b); 
        red dotted lines indicate fits with multiple Lorentzian peaks with slowly varying background.}
	\label{g1g2}
\end{figure}

Eq. (\ref{Andreev}) in combination  with the Lorentzian transmission probability of Eq. (\ref{tau_res})
has been demonstrated to reproduce accurately the shape of the conductance peaks in carbon nanotube quantum dots\cite{Gramich2015,Gramich2017}.
We find that this model is also working quite well for our graphene quantum dots.  
In Fig. \ref{g1g2} (a,b) we display the zero-bias conductance of the left and right quantum dot, measured at 
$T_L=T_R=T_S=90$ mK, as a function of the side gate voltages $V_{sg,L}$, $V_{sg,R}$.
Inverting the expression (\ref{Andreev}) for zero bias conductance, we obtain normal state transmission probabilities in the form
\begin{eqnarray}
\tau_j(0,V_{sg,j}) = \frac{2\sqrt{g_j(V_{sg,j})}}{2+\sqrt{g_j(V_{sg,j})}}.
\label{Tj}
\end{eqnarray}
Applying this transformation to the experimental data sets depicted in Figs. \ref{g1g2} (a,b), 
we get the normal state transmission probabilities at zero energy $\tau_j(0,V_{sg,j})$. The latter are plotted in Figs. \ref{g1g2} (c,d) with blue lines.
In the same figures, by red dotted lines we also show the fits of the transmissions $\tau_j(0,V_{sg,j})$ with multiple Lorentzian peaks in accordance
with the predictions of the resonance tunneling model (Eq. (\ref{tau_res})). The quality of these fits is rather good, especially 
for the left quantum dot.  We conclude, therefore, that the resonance tunneling model describes the quantum dots well.

\begin{figure}[!ht]
		\includegraphics[width=13cm]{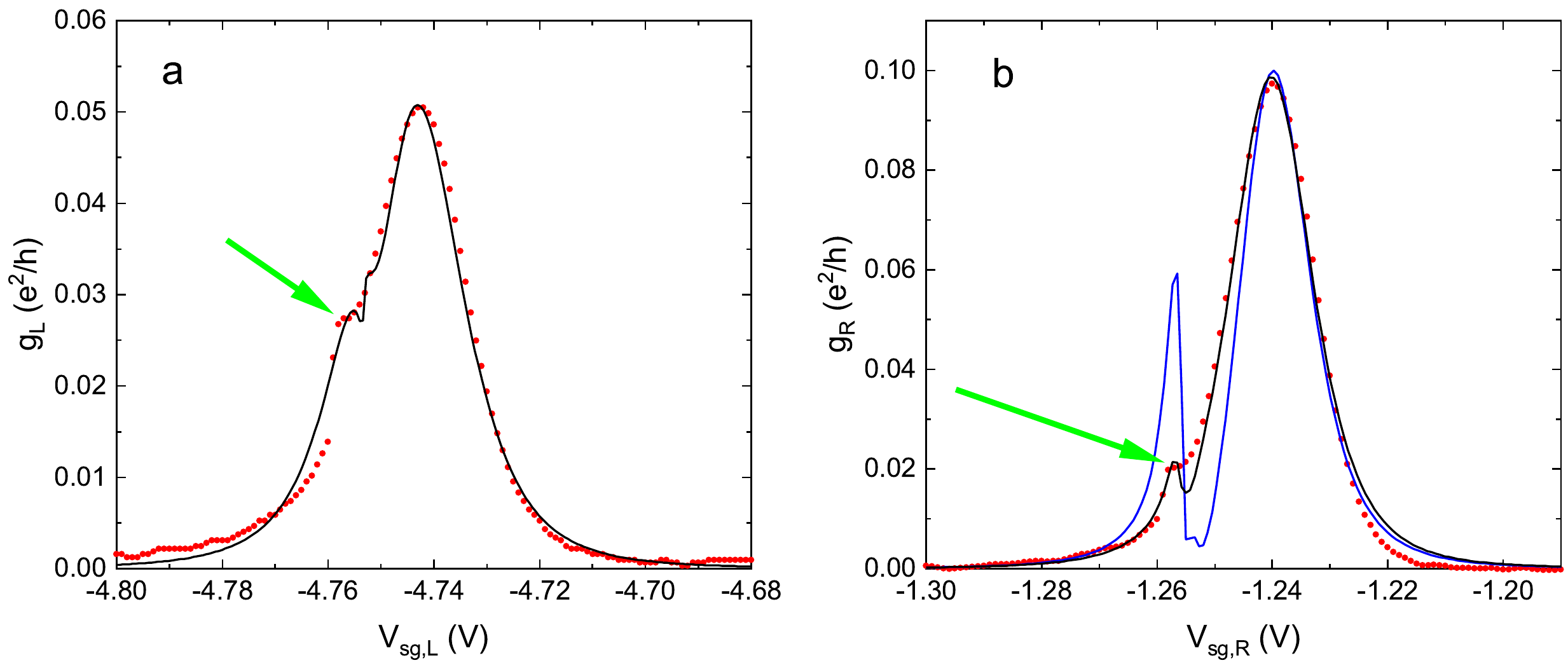} \\
	\caption{Fits of the two selected zero bias conductance peaks of the left (a) and the right (b) quantum dots, measured at $T=90$ mK,  with 
      Eq. (\ref{Ij}), in which the transmission probabilities have
     the Fano resonance form (\ref{tau_Fano}). Green arrows indicate weak additional peaks not captured by the simple resonant
      tunneling model (\ref{tau_res}).
(a)	One of the conductance peaks of the left dot; red dots show experimental data, black is the fit to Eqs. (\ref{tau_Fano},\ref{Ij}) with the parameters  
$\gamma_L=98.3$ $\mu$eV, $\Gamma_L=6.04$ $\mu$eV, $t_L=10$ $\mu$eV, $\varepsilon_L-\varepsilon'_L=24$ $\mu$eV.
(b) Conductance peak of the right dot; red dots -- experiment, black line is the best fit to Eqs. (\ref{tau_Fano},\ref{Ij}) with the parameters
$\gamma_R=260.2$ $\mu$eV, $\Gamma_R=20.8$ $\mu$eV $t_R=22$ $\mu$eV, $\varepsilon_R-\varepsilon'_R=155$ $\mu$eV,
blue line shows the theoretical curve with the parameters used in Fig. 4 of the main text, namely,
$\gamma_R=252.4$ $\mu$eV, $\Gamma_R=20.4$ $\mu$eV $t_R=55$ $\mu$eV, $\varepsilon_R-\varepsilon'_R=120$ $\mu$eV. 
The last set of parameters is used in Fig. 4 of the main text, in which the non-local contributions to the current are plotted.}
	\label{gLgR_fit}
\end{figure}

A closer look at some of the conductance peaks reveals deviations from the form predicted by Eq. (\ref{Andreev}) with the Lorentzian
transmission probabilities (\ref{tau_res}). 
In Fig. \ref{gLgR_fit} we plot two representative peaks, one for each dot, and indicate by the green arrows the
features not captured using the resonant tunneling model. These features may be caused by various reasons. For example, they may result from 
Coulomb interaction, inelastic relaxation processes, or Kondo effect, etc.
They may also arise from overlapping of the relevant energy levels. 
Here we fit these features using the Fano resonance model (\ref{tau_Fano}), in which a conductance peak splits into two closely lying peaks
with different heights and widths. We believe that this is the most plausible explanation, although we cannot fully exclude other options.
This uncertainty, however, does not affect the expression for the non-local current given below in Eq. (\ref{I12}) and in the main text. 
Indeed, the latter contains only the effective transmission probabilities $\tau_j$, which can be inferred from the conductance data, and, 
therefore, are not very sensitive to a specific model describing the features of $\tau_j(E,V_{sg,j})$ and its parameter dependence.   
With the Fano resonance model, we have fitted the conductance peak of the left quantum dot quite accurately, see Fig \ref{gLgR_fit}a. 
However, for the right dot we have adopted a more pronounced Fano resonance than the conductance data would suggest (Fig \ref{gLgR_fit}b) 
in order to reproduce extra sign changes of  the non-local thermal current in the vicinity of this conductance peak 
(cf. Fig. 4 in the main text). 
  
Having determined the zero-energy transmission probabilities of the quantum dots, $\tau_j(0,V_{sg,j})$, as functions of the gate voltages, 
we make the replacement $V_{gj}\to V_{gj} + E/\alpha_j$, where $E$ is the energy of an electron, and in this way recover the full  
energy dependence of the transmission probabilities. Then, we use the obtained energy-dependent transmission probability $\tau_L(E,V_{sg,L})$
to calculate theoretical thermoelectric currents, based on Eq. (\ref{Ij}), for different heating voltages.
The result, displayed in Fig. \ref{g1g2} of the main text, agrees rather well with the experimental findings apart for the magnitude of the current. We would like to emphasize that, in this model, only quasiparticles with the energies $|E|>\Delta$ contribute to the thermal current.
That is why the predicted values of the current at low heating powers, $I_j\propto \exp(-\Delta/T_j)$, are very small.
In the experiment, however, we observe significant thermal currents even at the lowest heating voltages  1 mV $< V_{h} <$ 5 mV.
This observation may be explained, for instance, by significant heating of the superconducting lead even at small heating voltages. Alternatively,
it may be a result of the induced thermal voltages, which can be explained by the incoherent tunneling model discussed in Sec. \ref{incoherent}.
  
For completeness, we provide the low temperature expressions for the local thermal currents, which 
follow from the general expression of Eq. (\ref{Ij}) for $V_L=V_R=0$ and in the limit $T_L,T_R,T_S\ll\Delta/k_B$,
\begin{eqnarray}
I_L^{\rm loc} &=& \frac{e\sqrt{2\pi\Delta}}{h}\left[\tau_L(\Delta)-\tau_L(-\Delta)\right]
\left[ \sqrt{k_BT_L} \, e^{-\Delta/k_BT_L} - \sqrt{k_BT_S} \, e^{-\Delta/k_BT_S} \right],
\nonumber\\
I_R^{\rm loc} &=& \frac{e\sqrt{2\pi\Delta}}{h}\left[\tau_R(\Delta)-\tau_R(-\Delta)\right]
\left[ \sqrt{k_BT_R} \, e^{-\Delta/k_BT_R} - \sqrt{k_BT_S} \, e^{-\Delta/k_BT_S} \right].
\end{eqnarray}

\subsection{Non-local contributions to the thermal current}

The main focus of our study is Cooper pair splitting (CPS).
It results from the crossed Andreev reflection process, in which a single Cooper pair splits into two electrons
injected into different quantum dots. It has been shown (see e.g. Refs. \onlinecite{Feinberg2000,Recher2001,Golubev2009,Golubev2019})
that the probability of such a process is proportional to the product of the transmission probabilities of the two quantum dots, $\tau_{L,R}(E)$,
and the effective transmission probability of the superconducting electrode $\tau_S$,
\begin{eqnarray}
\tau_{\rm CPS}(E)=\tau_L(E)\tau_S \tau_R(-E).
\label{CPS}
\end{eqnarray}   
Strictly speaking, Refs. \onlinecite{Feinberg2000,Recher2001,Golubev2009,Golubev2019} deal with metallic normal leads, in which the transmission
probabilities $\tau_L,\tau_R$ do not depend on the electron energy $E$. Correct energy dependence of these transmissions can be obtained
by comparing the crossed Andreev reflection probability of Eq. (\ref{CPS}) with the probability of ordinary Andreev reflection, in which
both electrons are emitted in to the same quantum dot. The latter probability appears in Eq. (\ref{subgap}) for the Andreev current
through a low transmission quantum dot and contains the product $\tau_j(E)\tau_j(-E)$, in which an electron and a reflected hole have opposite energies.
Clearly, the same rule should apply to the crossed Andreev reflection, and in this way one arrives at the symmetry of Eq. (\ref{CPS}). 
The effective transmission probability of the superconductor, $\tau_S$,
does not depend on the energy $E$ if the quantum dots are placed close to each other and the distance between them is less than the coherence length
of the superconductor. This condition is approximately satisfied in our experimental configuration.
The parameter $\tau_S$ has the meaning of the probability for an electron emitted from one of the dots to reach the other dot
instead of flying away into the bulk of the superconductor. 
From the theory of disordered superconductors based on the Usadel equation one formally finds \cite{Golubev2009}
\begin{eqnarray}
\tau_S = \frac{4\pi\hbar R_S}{e^2},
\label{tau_S}
\end{eqnarray}
where $R_S$ is the resistance of the superconducting lead in the normal state. Eq. (\ref{tau_S}) is formally valid if the
electron mean free path $l_e$ is much smaller than the distance between the dots.
In practice, the value of $\tau_S$ is sensitive to
the sample geometry, granularity of the superconductor, quality of the contacts between the superconductor and the quantum dots, etc.
Therefore, we treat it as a fit parameter. We find that $\tau_S=0.1$ provides good fit of our data.

It is well known that in addition to the Cooper pair splitting process, elastic cotunneling also contributes to the non-local transport.
The probability of this process is given by the product \cite{Feinberg2000,Recher2001,Golubev2009,Golubev2019}
\begin{eqnarray}
\tau_{\rm EC}(E)=\tau_L(E)\tau_S \tau_R(E),
\label{EC}
\end{eqnarray}    
which differs from $\tau_{\rm CPS}$ only by the replacement $E\to -E$ in the argument of $\tau_R$.
The approximate expressions in Eqs. (\ref{CPS},\ref{EC}) are valid provided  $\tau_{\rm CPS},\tau_{\rm EC}\ll 1$. This condition is well satisfied in our
experiment. 

The currents flowing through the quantum dots are given by the sum of  local and  non-local contributions,
\begin{eqnarray}
I_L = I_L^{\rm loc} + \Delta I_L^{\rm nl},\;\;\; I_R = I_R^{\rm loc} + \Delta I_R^{\rm nl},
\end{eqnarray}
where the local currents are given by Eq. (\ref{Ij}), and the non-local corrections
in the low temperature limit $k_BT_L,k_BT_R\ll \Delta$ and zero voltage drops across the dots, $V_j=0$, are expressed as
\begin{eqnarray}
\Delta I_L^{\rm nl} &=& \frac{e\tau_S}{h} \int dE \, \tau_L(E)\big[ \tau_R(E)+\tau_R(-E) \big]
\left[\frac{1}{1+e^{E/k_BT_L}}-\frac{1}{1+e^{E/k_BT_R}}\right], \;\;\;
\nonumber\\
\Delta I_R^{\rm nl} &=& -\frac{e\tau_S}{h} \int dE \, \big[ \tau_L(E)+\tau_L(-E) \big]\tau_R(E)
\left[\frac{1}{1+e^{E/k_BT_L}}-\frac{1}{1+e^{E/k_BT_R}}\right].
\label{I12}
\end{eqnarray}
One can also define the linear combinations of the non-local currents, which are determined solely by Cooper pair splitting or
elastic cotunneling,
\begin{eqnarray}
\Delta I_{\rm CPS} &=& \Delta I_L^{\rm nl} + \Delta I_R^{\rm nl} 
= \frac{2e}{h}\int dE\,\tau_{\rm CPS}(E) \left[\frac{1}{1+e^{E/k_BT_L}}-\frac{1}{1+e^{E/k_BT_R}}\right],
\nonumber\\
\Delta I_{\rm EC} &=& \Delta I_L^{\rm nl} - \Delta I_R^{\rm nl} 
= \frac{2e}{h}\int dE\,\tau_{\rm EC}(E) \left[\frac{1}{1+e^{E/k_BT_L}}-\frac{1}{1+e^{E/k_BT_R}}\right].
\label{I_CPS_EC}
\end{eqnarray}  
These expressions presented and discussed in the main text.

In order to understand the overall behaviour of the non-local currents, we again consider the low temperature limit 
$k_BT_L,k_BT_R\ll \Delta,\gamma_{j,n}+\Gamma_{j,n}$. In this limit one can derive the expressions analogous to Mott's formula (\ref{Ij_app}),
\begin{eqnarray}
\Delta I_L^{\rm nl} &=& \frac{\pi^2}{3} \frac{ek_B^2}{h} \frac{\partial\tau_L(0,V_{sg,L})}{\partial E} \tau_S \tau_R(0,V_{sg,R})(T_L^2-T_S^2)
\nonumber\\
&=& - \frac{\pi^2}{3} \frac{ek_B^2}{a_L h} \tau_S \frac{\partial g_L(0,V_{sg,L})}{\partial V_{sg,L}}  
\frac{4\sqrt{g_R(V_{sg,R})}}{\sqrt{g_L(V_{sg,L})}\left(2+\sqrt{g_L(V_{sg,L})}\right)^2\left(2+\sqrt{g_R(V_{sg,R})}\right)} (T_L^2-T_S^2),
\label{dIL}
\end{eqnarray}
\begin{eqnarray}
\Delta I_R^{\rm nl} &=& \frac{\pi^2}{3} \frac{ek_B^2}{h} \tau_L(0,V_{sg,L}) \tau_S  \frac{\partial\tau_R(0,V_{sg,R})}{\partial E} (T_R^2-T_S^2)
\nonumber\\
&=& - \frac{\pi^2}{3} \frac{ek_B^2}{a_R h} \tau_S \frac{\partial g_R(0,V_{sg,R})}{\partial V_{sg,R}}  
\frac{4\sqrt{g_L(V_{sg,L})}}{\sqrt{g_R(V_{sg,R})}\left(2+\sqrt{g_L(V_{sg,L})}\right)\left(2+\sqrt{g_R(V_{sg,R})}\right)^2} (T_R^2-T_S^2).
\label{dIR}
\end{eqnarray}
Eq. (\ref{dIL}), for example, shows that the non-local contribution to the thermal current of the left quantum dot should change sign 
at the values of the gate voltage $V_{sg,L}$ corresponding to the positions of the conductance peaks and the minima of
the conductance valleys of the left dot.
At the same time, no sign change in $\Delta I_L^{\rm nl}$ occurs if one varies the gate potential of the right dot $V_{sg,R}$. 
The current $\Delta I_L^{\rm nl}$ depends on the right gate potential $V_{sg,R}$ roughly in the same way as the conductance of the right dot $g_R(V_{sg,R})$ does.
Thus, in order to restore the sign pattern and relative magnitudes of the thermal currents one can use a simple rule
$\Delta I_L^{\rm nl} \propto (dg_L/dV_{sg,L})g_R$, $\Delta I_R^{\rm nl} \propto g_L(dg_R/dV_{sg,R})$. 
Such Mott-type behaviour of the non-local currents agrees well with gate voltage dependence observed in the experiment.

\clearpage
\setcounter{subsection}{0}
\setcounter{subsubsection}{0}
\section*{Note 5: Theoretical modeling: Incoherent transport in a N-dot-S-dot-N system. }\label{incoherent}
\begin{figure*}[h]
    \noindent\centering{
    \includegraphics[width=150mm]{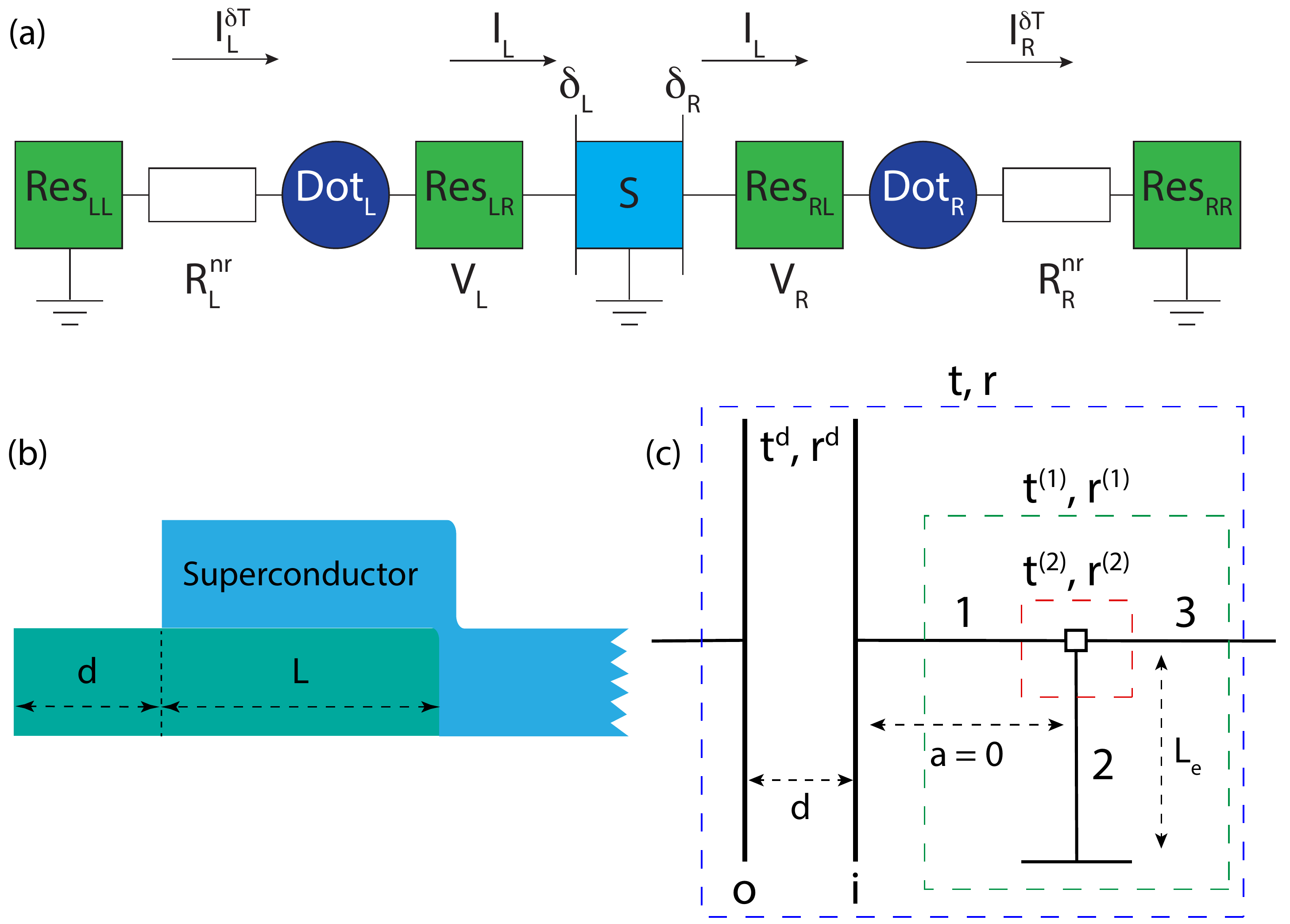}
    }
    \caption{(a) Noncoherent model of the experimental device.
    Res$_\text{LL}$, Res$_\text{LR}$, Res$_\text{RL}$, and Res$_\text{RR}$ are the electron-reservoirs, Dot$_\text{L(R)}$ is the left (right) quantum dot, $R^\text{nr}_L$ and $R^\text{nr}_R$ are the additional resistors, S is the superconducting region, $\delta_\text{L(R)}$ is $\delta$-barrier on the left (right) NS-interface.
    Res$_\text{LL}$, Res$_\text{RR}$ and S are grounded.
    (b) Side view (schematic) of the graphene-aluminum (superconductor) junction.
    The patterned graphene piece has a characteristic size of $\sim300$\,nm and constitutes a quantum dot.
    Here $L$ and $d$ are the lengths of the graphene parts which, respectively, are either overlapping or extending beyond the superconductor.
    (c) Schematics of the theoretical model accounting for the effect of the additional Fano resonance structure.     
    The quantum dot is modelled by two $\delta$-function potentials (the distance between barriers is $d$) connected to a fork with a stub of length $L_e\sim L^2/\lambda$.
    The transmission and reflection amplitudes of the beam splitter (blank square) are denoted by $t^{(2)}$ and $r^{(2)}$; the amplitudes incorporating the internal features of the fork and the stub are denoted by $t^{(1)}$ and $r^{(1)}$.
    We set the distance $a$ between the double barrier and the beam splitter equal to zero.}
    \label{noncoherent_scheme}
\end{figure*}

The model described in Note 4 is based on the assumption that the transport in the Dot-S-Dot system is fully coherent.
In reality, the system may be subject to dephasing and relaxation.
To demonstrate how the decoherence processes may affect the experiment, we present and analyze an alternative theoretical model based on the scattering matrix approach.
This model is predicated on the assumption that the system may be split into the coherent subsystems which, in turn, are joined incoherently
as connected circuit elements. For this reason, thermal gradients induce both electric currents and voltage drops on the dots. That is why,  
in contrast to the coherent transport model, the incoherent one predicts significant value of the local thermal currents at low temperatures, 
where the quasiparticles in the superconducting lead disappear, and, thus, it may explain the experimental observations in this regime.  
However, our analysis shows that decoherence has little effect on the measured non-local phenomena.

To account for possible decoherence, we assume that the schematics includes additional reservoirs (Res$_\text{LR}$ and Res$_\text{RL}$) between the superconductor and both the left and the right dot as depicted in Fig.\,\ref{noncoherent_scheme}(a).
We model the whole structure as a one-dimensional conducting structure with ballistic motion of electrons in the superconductor
and in the quantum dots. The superconductor (S) is separated by $\delta$-barriers ($\delta_\text{L}$ and $\delta_\text{R}$) from the adjacent reservoirs.
The transmission probabilities of the quantum dots (Dot$_\text{L}$ and Dot$_\text{R}$) may exhibit  Fano resonances.
In order to model such resonances, the dots are assumed to be composed of two elements --- Fabry-P\'{e}rot double barrier structure and a stub. 
Finally, we assume that the leftmost and rightmost reservoirs (Res$_\text{LL}$ and Res$_\text{RR}$) along with the superconductor are grounded.
\label{incoherent_Fig}

\subsection{Local electric current}
Let us fist discuss the characteristic features of the experimental plots showing the dependence of the local thermoelectric current on the gate voltage in the quantum dot.
For the present model we assume that the local thermoelectricity in the left (right) side of the structure is governed by the difference between the population distributions in Res$_\text{LL}$ and Res$_\text{LR}$ (Res$_\text{RL}$ and Res$_\text{RR}$) due to the temperature gradient.
For simplicity, we will consider only the subgap regime.

As mentioned in the main paper, we can explain the extraordinary behavior of the local thermoelectric current (i.e., the appearance of a secondary extremum on one side of the main inflexion point) by the Fano resonant effect emerging in the connection between the dot and the superconductor.
The arrangement of the experimental setting (as shown in Fig.\,\ref{noncoherent_scheme}(b)) suggests that it can be modeled by the scheme depicted in Fig.\,\ref{noncoherent_scheme}(c). 
%where the graphene quantum dot is modelled by a double $\delta$-function barrier and the connection to the superconductor lies in between the barriers.
%The same model can be reformulated in a more convenient way as shown in Fig.\,\ref{scheme}(d), where above all we add an additional delta barrier at distance $d$ from the left one (in comparing to the equivalent scheme in Fig.\,\ref{scheme}(c), one should imply that the scattering on this additional barrier is described by the scattering matrix of the connection point denoted by blank circle). 
Here the interface between the superconductor and the double delta barrier (the distance between barriers is $d$) is considered to have a fork structure with one of its contacts being a stub of effective length $L_e$. 
$L_e$ is much larger than $d$ which ensures the interaction between the discreet spectrum due to the stub and the semi-continuum in the double delta barrier (which encompasses a large level width).
Drawing a parallel with the schematics in Fig.\,\ref{noncoherent_scheme}(b), the stub relates to the part of graphene overlapping with the superconductor, which has length $L$.
The length of the stub $L_e$, however, should not necessarily be equal to $L$: 
since the mean free path $\lambda$ of the particles is smaller than $L$, in reality they may undergo multiple reflections inside this part of the structure. Then
\begin{equation}
    L^2\sim D t_{t},
\end{equation}
where $D \sim v_F \lambda$ is the diffusion coefficient and $t_{t}$ is the time that the particle spends in the stub.
Accordingly, the effective length of the stub $L_e \sim v_F t_{t}\sim L^2/\lambda$ used in the model can exceed the typical size of the dot $L\simeq300$\,nm.
We assume that the barrier in the end of the stub completely blocks the propagation of the particles, i.e., the probability density for finding the particle beyond the barrier is zero.
For clarity, we separate the right delta barrier from the beam splitter, but imply that the distance $a$ between them is zero.
If $L_e$ is finite, the appearance of Fano resonances affects the transparency between left and right terminals changing the behaviour of the thermoelectric current.

% %
% \begin{figure}[t]
%     \noindent\centering{
%     \includegraphics[width=80mm]{theory.eps}
%     }
%     \caption{
%     Thermoelectric current as function of the resonance position $E_D$ in the absence of the additional Fano resonance structure; the transparency function is given by Eq. (\ref{lorenz}). 
%     }
%     \label{theory}
% \end{figure}
% %
%
% \begin{figure}[t]
%     \noindent\centering{
%     \includegraphics[width=90mm]{Fig_S12.png}
%     }
%     \caption{
%     (a) Side view (schematic) of the graphene-aluminum (superconductor) junction.
%     The patterned graphene piece has a characteristic size of $\sim300$\,nm and constitutes a quantum dot.
%     Here $L$ and $d$ are the lengths of the graphene parts which, respectively, are either overlapping or extending beyond the superconductor.
%     %
%     (b) Schematics of the theoretical model accounting for the effect of the additional Fano resonance structure.     
%     The quantum dot is modelled by two $\delta$-function potentials (the distance between barriers is $d$) connected to a fork with a stub of length $L_e\sim L^2/\lambda$.
%     The transmission and reflection amplitudes of the beam splitter (blank square) are denoted by $t^{(2)}$ and $r^{(2)}$; the amplitudes incorporating the internal features of the fork and the stub are denoted by $t^{(1)}$ and $r^{(1)}$.
%     We set the distance $a$ between the double barrier and the beam splitter equal to zero.
%     %
%     }
%     \label{loc_dot_scheme}
% \end{figure}
% %
%
\begin{figure}[t]
    \noindent\centering{
    \includegraphics[width=90mm]{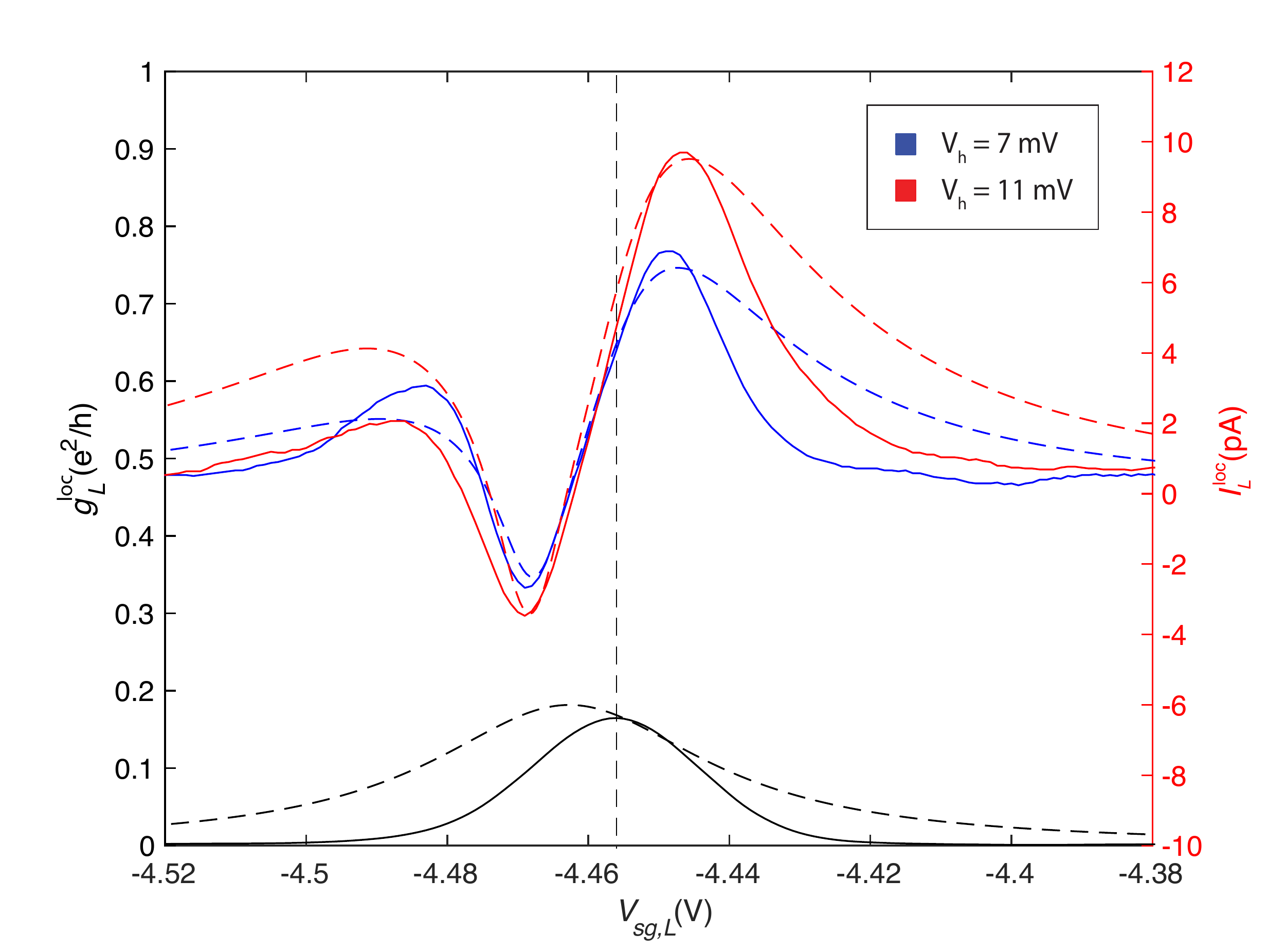}
    }
    \caption{
    Uppermost curves: fits (dashed) of the experimentally measured local thermoelectric currents (solid) at two different heating voltages $V_h$ (7 and 11 mV).
    Lower traces: experimentally measured (solid) and theoretically modelled (dashed) conductance.
    The parameters are $t_{13}^{(2)}=0.01\, e^{-0.03i}$, $t_{23}^{(2)}=0.55$, $Z_o=5$, $Z_i=5$, $R=5\frac{h}{e^2}$, $2 k_F L_e=2.33+2\pi n$ ($n$ is integer), $d=2.0$\,nm, $L_{e} = 1810.0$\,nm (we assume that the Fermi velocity in graphene is $v_F=10^6 m/s$).
    %$\phi=2.55$. 
    We take $T_{LL}=(0.09+9.1\, V_h^{0.70})$\,K and $T_{LR}=(0.09+8.67\, V_h^{0.70})$\,K.
    }
    \label{fits}
\end{figure}
Next we derive the transparency function which would account for the occurrence of this additional Fano scattering in the dot structure.
\\
\textit{Double delta barrier.--} The amplitudes of the transmission and reflection (from the inner delta barrier) associated with the double barrier can be written in the form
\begin{gather}
    \label{FP_t}
    t^d = t_i t_o e^{i k_d d}
    /(1-r_i r_o e^{2 i k_d d});
    \\
    \label{FP_r}
    r_{i}^{d} = r_i + r_o t_i^2 e^{2 i k_d d}
    /(1-r_i r_o e^{2 i k_d d});
\end{gather}
where $t_{i(o)}=1/(1+iZ_{i(o)})$ and $r_{i(o)}=-iZ_{i(o)}/(1+iZ_{i(o)})$ are the transmission and reflection amplitudes of the inner (outer) delta barrier (expressed in terms of strengths $Z_{i}$ and $Z_{o}$), $d$ is the distance between delta barriers and
\begin{equation}
    {k_d= k_F+\frac{E-e V_g}{\hbar v_F}}
\end{equation}
is the wave vector of the particle inside the double barrier (here, $V_g$ is the voltage on the gate).
\\
\textit{Fork.--} Supposing that the wave function of the particles goes to zero on the boundary in the stub and the stub is grounded (i.e., it is not affected by the gate voltage on the dot), we may obtain the transmission and reflection amplitudes $t^{(1)}$, $r^{(1)}$ on the contact 1-3 (incorporating the internal features, i.e., the stub, see {Fig.\,\ref{noncoherent_scheme}(c)}):
\begin{gather}
    t^{(1)}_{13}=t^{(2)}_{13}-\frac{t^{(2)}_{12}t^{(2)}_{23}}{\exp{(-2ikL_e)}+r^{(2)}_{22}},\\
    r^{(1)}_{33}=r^{(2)}_{33}-\frac{t^{(2)}_{32}t^{(2)}_{23}}{\exp{(-2ikL_e)}+r^{(2)}_{22}},\\
    r^{(1)}_{11}=r^{(2)}_{11}-\frac{t^{(2)}_{12}t^{(2)}_{21}}{\exp{(-2ikL_e)}+r^{(2)}_{22}},
\end{gather}
where index $(2)$ denotes the transmission and reflection amplitudes of the beam splitter, and $L_e$ is the length of the stub;
we assume that the wave vector {$k = k_F+\frac{E}{\hbar v_F}$} in the stub does not depend on the gate voltage. 

For the numerical calculations we will parametrize the beam splitter's scattering matrix with only two parameters:
the elements of the matrix can be expressed in terms of the transmission probabilities in contacts 1-3 and 2-3:
\begin{gather}
    \label{param}
    \tau_{12}^{(2)}=\tau_{13}^{(2)} \tau_{23}^{(2)}\left[2-\tau_{\Sigma}^{(2)} + 2\left(1-\tau_{\Sigma}^{(2)}\right)^{1 / 2}\right] / (\tau_{\Sigma}^{(2)})^{2},\\
    r_{i i}^{(2)}=t_{j k}^{(2) *} t_{i j}^{(2)} t_{i k}^{(2)}\left[(\tau_{j k}^{(2)})^{-1}-(\tau_{i j}^{(2)})^{-1}-(\tau_{i k}^{(2)})^{-1}\right] / 2,
\end{gather}
where $\tau_{\Sigma}^{(2)} \equiv \tau_{13}^{(2)}+\tau_{23}^{(2)}, \text { and } \tau_{i j}^{(2)}=\left|t_{i j}^{(2)}\right|^{2}$.
\\
\textit{Whole structure.--} The final transmission probability $\tau=|t|^2$ of the dot can be calculated by substituting Eqs.\,(\ref{FP_t}-\ref{param}) into the formula
\begin{equation}
\label{new_trans}
    t=\frac{t^d t^{(1)}_{13}}{1-r^{(1)}_{11} r^d_i}.
\end{equation}
The scattering on the outer (O) and inner (I) points of the structure is described by the reflection amplitudes
\begin{gather}
    r_{I} = r^{(1)}_{33}+\frac{r^d_i (t_{13}^{(1)})^2}{1-r^{(1)}_{11} r^d_i};\\
    r_{O} = r^{d}_{o}+\frac{r^{(1)}_{11} (t^{d})^2}{1-r^{(1)}_{11} r^d_i}.
\end{gather}

In the experiment, the local conductance and current are measured in the network of the dot along with the graphene nanoribbon (represented by $R^{\rm nr}_{L(R)}$ in Fig.\,\ref{noncoherent_scheme}(a)) and the N-$\delta$-S boundary.
For the theoretical model, we should assume that the nanoribbon, together with the N-$\delta$-S boundary, have a constant electrical resistance $R$. 
This allows us to write the following equation for the local current through the left dot (the transparency of the dot $\tau_L$ is calculated as discussed above):
\begin{equation}
    I_L^{\rm loc}=\frac{2e}{h}\int dE \Bigg[\frac{1}{\exp{\frac{E+e(-V+I_L^{\rm loc}R)\alpha}{k_B T_{LL}}}+1}-\frac{1}{\exp{\frac{E+e(-V+I_L^{\rm loc}R)(\alpha-1)}{k_B T_{LR}}}+1}\Bigg]\,\tau_L(E),
\end{equation}
where $V$ denotes the external voltage applied to the system (we will need $V$ later to calculate the differential conductance), $T_{LL(LR)}$ is the temperature of Res$_{LL(LR)}$ and $\alpha$ ($0\leq\alpha\leq1$) describes the voltage distribution across the dot. 
Since $\delta T_L=T_{LL}-T_{LR}\ll T_{LR}$ and $eI_L^{\rm loc}R \ll k_B T_{LR}$ (as one will be able to see later), we obtain
\begin{equation}
    I_L^{\rm loc}=\frac{2e}{h}\int dE \frac{\partial f}{\partial E}[-eV+eI_L^{\rm loc}R-E \delta T_L/T_{LR}]\,\tau_L(E),
\end{equation}
whence
\begin{equation}
    I_L^{\rm loc}=\frac{-\frac{2e}{h}\int dE \frac{\partial f}{\partial E}[eV+E \delta T_L/T_{LR}]\,\tau_L (E)}{1 - R\cdot\frac{2e^2}{h}\int dE \frac{\partial f}{\partial E}\,\tau_L (E)}.
\end{equation}
The parameters of our model are such that the dependence of $\tau_L (E)$ on $E$ is relatively weak. 
We can therefore write the following equations for the current $I_L^{\rm loc}$ ($V=0$) and the differential conductance $g_L^{\rm loc}=\frac{\partial I_L^{\rm loc}}{\partial V}$:
\begin{gather}
    % I=-\frac{2 e}{h}\frac{\int dE\,E\,\frac{\partial f}{\partial E}\, \mathcal{T}(E)\, \delta T/T_R}{1+\frac{2e^2}{h} \mathcal{T}(0) R}
    \label{final}
    I_L^{\rm loc}=\frac{2 \pi^2 e k_B^2 T_{LR} \delta T_L}{3h}\frac{{\partial \tau_L}/{\partial E}(0)}{1+\frac{2e^2}{h} \tau_L(0) R};
    \\
    g_L^{\rm loc} = \frac{2 e^2}{h}\frac{\tau_L(0)}{1+\frac{2e^2}{h} \tau_L(0) R}.
\end{gather}

Fig.\,\ref{fits} depicts experimental fits for different heating voltages $V_h$ obtained using Eqs.\,(\ref{param}), (\ref{new_trans}) and (\ref{final}); we put temperatures $T_{LL}=(0.09+9.1\, V_h^{0.70})$\,K and $T_{LR}=(0.09+8.67\, V_h^{0.70})$\,K; other parameters are given in the caption.
One can see that, for small chosen values of $\tau_{13}^{(2)}$, the behaviour of the thermoelectric current predicted theoretically compares well with the experimental data.
Moreover, we should note that the parameters of the fits are such that the length $L_e=1810$\,nm comply with our prediction of $\sim L^2/\lambda \sim 2000$\,nm given that $L \simeq 300$\,nm and $\lambda\sim 30$\,nm.
Thus, the above 
%coherent 
model, accounting for quasi-particles in the superconductor at elevated temperatures, allows for a faithful theoretical description of the measured data.
However, the model does not reproduce the experimental observation that the secondary features of the curves tend to disappear at large heating voltages. 
% Such a phenomenon may for instance be explained by the temperature dependence of the amplitudes $t^{(2)}$, $r^{(2)}$ due to the proximity effect.

\subsection{Non-local electric current}
The non-local currents in this model are contributed  by two factors:
\begin{itemize}
    \item The voltage difference between Res$_\text{LR}$ and Res$_\text{RL}$ which, in turn, is induced by the local thermoelectricity in the left and right leads. 
    For instance, the temperature difference between Res$_\text{LL}$ and Res$_\text{LR}$ generates a nonzero electric potential in Res$_\text{LR}$.
    \item The temperature gradient between Res$_\text{LR}$ and Res$_\text{RL}$.
\end{itemize}
Our numerical analysis shows that the latter factor does not play a significant role and can be neglected.
Physically, this is due to the fact that the transmission through the $\delta_\text{L}$-S-$\delta_\text{R}$ structure depends on the energy of the incident particles $E$ rather weakly compared to the transmission through the dots.
%Here, however, we should not pause to discuss this any further.

The left-to-right current in the left lead, which flows from Res$_{LR}$ to S and Res$_{RL}$ is given by\,\cite{Kirsanov2019} 
\begin{multline}
\label{IL}
I_{L}(T_{LR},\,T_{RL},\,V_{L},\,V_{R}) = \frac{2e}{h} \int dE\, \{-\mathcal{R}^{LL}_{eh}(E)\, [1 - f(E-eV_L,\,T_{LR})]-\mathcal{T}^{RL}_{eh}(E)\, [1 - f(E-eV_R,\,T_{RL})]\\
+[1 - \mathcal{R}^{LL}_{ee}(E)]\,f(E-eV_L,\,T_{LR})-\mathcal{T}^{RL}_{ee}(E)\, f(E-eV_R,\,T_{RL})\},
\end{multline}
where $T_{LR(RL)}$ and $V_{L(R)}$ are, respectively, the temperature and electric potential of Res$_{LR(RL)}$, 
%$f(E,\,T)=\frac{1}{\exp(\frac{E}{k_B T})+1}$ is the Fermi distribution function,
$\mathcal{R}^{LL}_{eh(ee)}(E)$ is the probability of Andreev (normal) reflection on the N-$\delta_L$-S interface, $\mathcal{T}^{RL}_{eh(ee)}$ is the probability of the Cooper pair splitting (elastic co-tunneling) with regard to the particles incident from the right lead (here the upper subscript indicates the direction of the particle motion; e.g., $RL$ means that the particle incident from the right lead is transmitted into the left one). 

The probabilities $\mathcal{T}^{LL}_{eh(ee)} = |\tilde{r}_{eh(ee)}|^2$ and $\mathcal{R}^{RL}_{eh(ee)} = |\tilde{t}_{eh(ee)}|^2$ are determined by the transmission and reflection amplitudes given by\,\cite{Sadovskyy2015}
\begin{gather}
    \tilde{t}_{eh}=t_L\,[t_{ee}\,r_{R}\,r_{eh} + r_{eh}\,r_{L}\,t_{hh}]\,t_R/\mathcal{D};\\
    \tilde{t}_{ee}=t_L\,[t_{ee}\,(1-t_{hh}^2\,r_{L}\,r_{R})+r_{eh}\,r_{L}\,t_{hh}\,r_{R}\,r_{he}]\,t_R/\mathcal{D};\\
    \tilde{r}_{eh}=t_{L}\,r_{eh}\,[1+(t_{ee}\,t_{hh}-r_{eh}\, r_{he})\,r_{R}\,r_{R}]\,t_{L} / \mathcal{D};\\
    \tilde{r}_{ee}= r_L + t_L\,[r_{eh}\,r_L\,r_{he} + t_{ee}\,r_R\,t_{ee}-(t_{ee}\,t_{hh}-r_{eh}\, r_{he})^2\,r_L\,r_R\,r_R]\,t_L/ \mathcal{D},
\end{gather}
where $t_{L(R)}$ and $r_{L(R)}$ are the transmission and reflection amplitudes of $\delta_{L(R)}$; $\mathcal{D}$ is determined by multiple reflections inside the $\delta_L$-S-$\delta_R$ structure:
\begin{equation}
    \mathcal{D}=1-t^2_{ee}\,r_{L}\,r_{R} - t^2_{hh}\,r_{L}\,r_{R} - r_{eh}\,r_{he}\,(r_{L}\,r_{L} + r_{R}\,r_{R})
    + (t_{ee}\,t_{hh}-r_{eh}\,r_{he})^2\,r_{L}\,r_{R}\,r_{L}\,r_{R};
\end{equation}
and $t_{ee(hh)}$ and $r_{ee(hh)}$ are the trasmission and reflection amplitudes of the superconducting part of the structure:
\begin{gather}
    t_{ee(hh)}=\frac{e^{\pm i p l_S}\sin{\alpha}}{\sin{(\alpha-i q l_S)}};\\
    r_{eh(he)}=\frac{\sinh{q l_S}}{i \sin{(\alpha-i q l_S)}}.
\end{gather}
Here $p^2-q^2=k_F^2$ and $2pq=(2m/\hbar^2)\Delta\sin{\alpha}$ with $\alpha=\arccos{(E/\Delta)}$; $l_S$ is the size of the superconducting region.

Bearing in mind that $I_{L}(T,\,T,\,0,\,0)=0$ and that the temperature difference between Res$_{LR}$ and Res$_{RL}$ has a small contribution to the thermoelectric current, from Eq.\,(\ref{IL}) we obtain
\begin{multline}
\label{ILaprox}
I_{L}(T_{LR},\,T_{RL},\,V_{L},\,V_{R}) \simeq \frac{2e^2}{h} \int dE\, \{-\mathcal{R}^{LL}_{eh}(E)\,\frac{\partial f}{\partial E}(E,\,T_{LR})\,V_L
-\mathcal{T}^{RL}_{eh}(E)\,\,\frac{\partial f}{\partial E}(E,\,T_{RL})\,V_R\\
-[1 - \mathcal{R}^{LL}_{ee}(E)]\,\frac{\partial f}{\partial E}(E,\,T_{LR})\,V_L + \mathcal{T}^{RL}_{ee}(E)\,\frac{\partial f}{\partial E}(E,\,T_{RL})\,V_R
\\\simeq \frac{2e^2}{h} [2\mathcal{R}^{LL}_{eh}(0)+2\mathcal{T}^{LR}_{eh}(0)]\,V_L + \frac{2e^2}{h} [\mathcal{T}^{LR}_{ee}(0)-\mathcal{T}^{LR}_{eh}(0)]\,V_L - \frac{2e^2}{h} [\mathcal{T}^{RL}_{ee}(0)-\mathcal{T}^{RL}_{eh}(0)]\,V_R\\
=G_L^{NS}\,V_L + G_L^{NSN}\,V_L - G_R^{NSN}\,V_R,
\end{multline}
where the quantities defined as
\begin{gather}
    G_{L(R)}^{NS}= \frac{2e^2}{h} [2\mathcal{R}^{LL(RR)}_{eh}(0)+2\mathcal{T}^{LR(RL)}_{eh}(0)],\\
    G_{L(R)}^{NSN}=\frac{2e^2}{h} [\mathcal{T}^{LR(RL)}_{ee}(0)-\mathcal{T}^{LR(RL)}_{eh}(0)]
\end{gather}
may be said to represent, respectively, N-$\delta$-S and N-$\delta$-S-$\delta$-N conductances.
Similarly, the left-to-right current in the right lead, flowing from S and Res$_{LR}$ to Res$_{RL}$ is given by
\begin{multline}
\label{IR}
I_{R}(T_{LR},\,T_{RL},\,V_{L},\,V_{R}) = -\frac{2e}{h} \int dE\, \{-\mathcal{R}^{RR}_{eh}(E)\, [1 - f(E-eV_R,\,T_{RL})]-\mathcal{T}^{LR}_{eh}(E)\, [1 - f(E-eV_L,\,T_{LR})]\\
+[1 - \mathcal{R}^{RR}_{ee}(E)]\,f(E-eV_R,\,T_{LR})-\mathcal{T}^{LR}_{ee}(E)\, f(E-eV_L,\,T_{LR})\}\\
\simeq -G_R^{NS}\,V_R + G_L^{NSN}\,V_L - G_R^{NSN}\,V_R.
\end{multline}

Assuming now that $\delta_L$ and $\delta_R$ are identical, we have $G_L^{NS}=G_R^{NS}\equiv G^{NS}$ and $G_L^{NSN}=G_R^{NSN}\equiv G^{NSN}$.
Thus, 
\begin{gather}
\label{currents}
    I_L = G^{NS}\,(V_L-0) + G^{NSN}\,(V_L-V_R);\\
    I_R = G^{NS}\,(0-V_R) + G^{NSN}\,(V_L-V_R).
\end{gather}
Using Kirchhoff's law for the currents at the intersection of Res$_{LR}$, Res$_{RL}$, and the superconductor, we obtain the following equations:
\begin{gather}
    I_L^{\delta T} + g_L\,(0-V_L) = G^{NS}\,(V_L-0) + G^{NSN}\,(V_L-V_R);\\
    I_R^{\delta T} + g_R\,(V_R-0) = G^{NS}\,(0-V_R) + G^{NSN}\,(V_L-V_R),
\end{gather}
where $g_{L(R)}$ is the conductance of Dot$_{L(R)}$ plus additional resistor (nanoribbon) $R^{\rm nr}_{L(R)}$, and $I_{L(R)}^{\delta T}$ is the locally induced thermoelectric current:
\begin{gather}
    g_{L(R)} = \left(\frac{1}{\frac{2e^2}{h}\tau_{L(R)}(0)}+R^{\rm nr}_{L(R)}\right)^{-1}=\frac{2 e^2}{h}\frac{\tau_{L(R)}(0)}{1+\frac{2e^2}{h} \tau_{L(R)}(0) R^{\rm nr}_{L(R)}};\\
    I_{L}^{\delta T}=\frac{2e}{h}\int dE \,\tau_{L}(E)\,(f(E,\,T_{LL})-f(E,\,T_{LR}));\\
    I_{R}^{\delta T}=\frac{2e}{h}\int dE \,\tau_{R}(E)\,(f(E,\,T_{RL})-f(E,\,T_{RR})).
\end{gather}
We thus obtain
\begin{gather}
    V_R=\frac{-I_R^{\delta T}+I_L^{\delta T}\,G^{NSN}/(g_L+G^{NS}+G^{NSN})}{g_R+G^{NS}+G^{NSN}-(G^{NSN})^2/(g_L+G^{NS}+G^{NSN})};\\
    V_L=\frac{I_L^{\delta T}+G^{NSN}\,V_R}{g_L+G^{NS}+G^{NSN}}.
\end{gather}
The currents can be found by substituting $V_L$ and $V_R$ into Eq.\,(\ref{currents}).
To select the non-local contribution from the $I_{L(R)}$, one has to subtract $\langle I_{L(R)}\rangle$ (averaging should be performed over different gate voltages on the right (left) dot) from $I_{L(R)}$:
\begin{equation}
     \Delta I_{L(R)}^{\rm nl} = I_{L(R)} - \langle I_{L(R)}\rangle.
\end{equation}
\begin{figure*}[h!]
    \noindent\centering{
    \includegraphics[width=90mm]{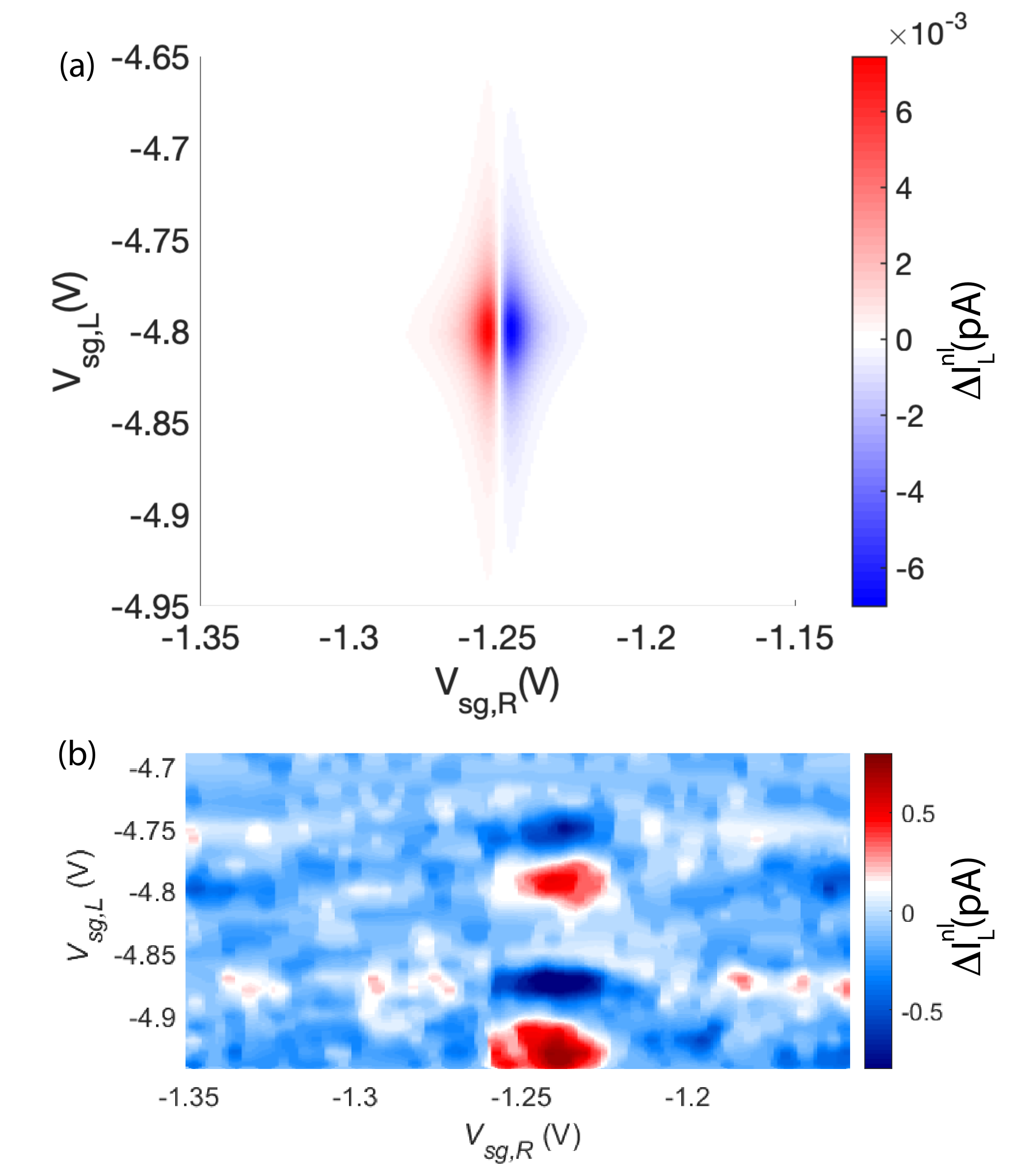}
    }
        \caption{Non-local current $\Delta I_{L}^{\rm nl}$ on the left side as function of the gate voltages $V_{sg,L}$ and $V_{sg,R}$ obtained (a) theoretically using the incoherent model and (b) experimentally.
        The parameters for (a) are such that the Fano resonance structure in the dots is absent: the parameters of both dots are
        $t_{13}^{(2)}=1$, $t_{23}^{(2)}=0$, $Z_o=5$, $Z_i=5$; 
        $R^\text{nr}_L=R^\text{nr}_R=5\frac{h}{e^2}$, $d=2.0$\,nm, the Fermi energy $E_{F} \gg \Delta$, $l_S=k_{F} \Delta /\left(2 E_{F}\right)$, $p l_S=0.9\pi +2\pi n$ ($n$ is integer);
        $t_L=t_R=1/(1+i)$, $r_L=r_R=-i/(1+i)$
        $T_{LL}=(0.09+9.1\, V_h^{0.70})$\,K, $T_{LR}=(0.09+8.8\, V_h^{0.70})$\,K, $T_{RL}=(0.09+8.8\, V_h^{0.70})$\,K, $T_{RR}=(0.09+8.4\, V_h^{0.70})$\,K.
        }
    \label{nonlocal}
\end{figure*}

As indicated by the calculations above, in the case of the incoherent system, the non-local electrical currents mostly originates from the electric potential difference between the intermediate reservoirs Res$_{LR}$ and Res$_{RL}$.
Therefore, in a strict sense, $\Delta I_{L(R)}^{\rm nl}$ is not thermoelectric current but rather it is produced by the locally induced thermoelectric voltages $V_L$ and $V_R$ on Res$_{LR}$ and Res$_{RL}$, which in this regard act as proxies.
However, the plots for $\Delta I_{L(R)}^{\rm nl}$ obtained from the present model are fundamentally different from the experimental data, which suggests that the observed non-local currents have a direct thermoelectric nature:
Fig.\,\ref{nonlocal} displays the (a) theoretically and (b) experimentally obtained non-local current $\Delta I_{L}^{\rm nl}$ on the left side as function of the gate voltages $V_{sg,L}$ and $V_{sg,R}$ in a simple situation where $\tau_L(E)$ and $\tau_R(E)$ have a Lorentzian form (i.e., the Fano resonance effects are absent).
Both plots are characterized by similar patterns, but one can easily observe that they have different orientations of their symmetry axes. 
This indicates that while the experimental setup can still be subject to decoherence, which may enable the above discussed mechanism for the local thermoelectricity, the non-local current is mostly determined by the coherent electrical transport.
Note that the exact values of $\Delta I_{L(R)}^{\rm nl}$ in Fig.\,\ref{nonlocal}(b) are not important since the purpose is rather to show the distinctive behaviour of the non-local electricity conditioned by the incoherent transport.

To conclude, the incoherent description accurately predicts the character of the local thermoelectricity at small temperatures. 
At the same time, at variance to effect of the local thermoelectricity, the non-local currents are dominantly determined by coherent electrical transport.

%\end{widetext}

\end{document}